\documentclass[preprint2,twocolumn]{aastex631}

\shorttitle{Mass outflow in {\fontfamily{qcr}\selectfont TCAF}}
\shortauthors{Mondal \& Chakrabarti}
\graphicspath{{./}{Figures/}}

\begin{document}

\title{Spectral signature of mass outflow in Two Component Advective Flow Paradigm}


\author{Santanu Mondal$^\ast$}
\affiliation{Indian Institute of Astrophysics, II Block, Koramangala, \\
Bangalore 560034, India}

\author{Sandip K. Chakrabarti}
\affiliation{Indian Centre for Space Physics, 43 Chalantika, \\
Garia Stn. Road, Kolkata 700084, India}

\email{santanuicsp@gmail.com; sandipchakrabarti9@gmail.com}

\begin{abstract}

Outflows are common in many astrophysical systems. In the Two Component Advective Flow ({\fontfamily{qcr}\selectfont TCAF})
paradigm which is essentially a generalized Bondi flow including rotation, viscosity and cooling effects, the outflow is
originated from the hot, puffed up, post-shock region at the inner edge of the accretion disk. We consider this region to be
the base of the jet carrying away matter with high velocity. In this paper, we study the spectral properties of
black holes using {\fontfamily{qcr}\selectfont TCAF} which includes also a jet  ({\fontfamily{qcr}\selectfont JeTCAF})
in the vertical direction of the disk plane. Soft photons from the Keplerian disk are up-scattered by the post-shock region
as well as by the base of the jet and are emitted as hard radiation. We also include the bulk motion Comptonization
effect by the diverging flow of jet. Our self-consistent accretion-ejection
solution shows how the spectrum from the base of the jet varies with accretion rates,
geometry of the flow and the collimation factor of the jet.
We apply the solution to a jetted candidate GS\,1354-64 to estimate its mass outflow rate and the geometric configuration
of the flow during 2015 outburst using {\it NuSTAR} observation. The estimated mass outflow to mass inflow rate is $0.12^{+0.02}_{-0.03}$.
From the model fitted accretion rates, shock compression ratio and the energy spectral index,
we identify the presence of hard and intermediate spectral states of the outburst. Our model fitted jet
collimation factor ($f_{\rm col}$) is found to be $0.47^{+0.09}_{-0.09}$.

\end{abstract}


\section{Introduction} \label{sec:intro}
Observations show that jets are common in active galactic nuclei (AGN), stellar mass
black holes and neutron stars \citep[e.g.,][and references therein]{Hjellming1995,Mirabel1995,MillerJones2012,Blandfordetal2019}. 
A proper understanding of the origin and acceleration of jets in compact objects is, however, lacking till date. 
Several models in the literature attempt to explain the origin, powering,
and collimation  of the jet. \citet{DsilvaChak1994} suggested that only the dominant 
toroidal field in the disk would be enough to accelerate and collimate the jets. This has been recently verified 
using numerical simulation by \citet{Garainetal2020}. 
\citet{Fender2010} from their observational studies did not find any evidence of
jet powered by the spin energy of the black holes and the authors speculated that this conclusion remained 
valid even for active galaxies and quasar. On the contrary, \citet{Narayan2012} found a correlation 
between jet power and black hole spin energy. Thus, the role of 
spin parameter in powering jet is still unclear and more studies must be done to reach a firm conclusion.
The hydrodynamic jets can also be collimated due to the strength of 
the jet-cocoon interaction and the collimation shock at the base of the jet \citep{Bromberg2011}.

Therefore, accretion-ejection around black hole systems has three components: 
the accretion disk, its static or dynamic corona, and 
the outflowing jet. It is well established that the standard disk \citep{ShakuraSunyaev1973} emits soft photons, 
up-scattered by the hot Compton cloud at the inner region of the disk and emit as hard radiation. The emergent radiation
may get down-scattered by the diverging outflows for both sub- and super-Eddington accretion 
\citep[][hereafter TS05]{Kawashimaetal2012,TitarchukShrader2005}. This effect changes the intrinsic continuum
spectra of accretion disks with an additional bump \citep{SunyaevTitarchuk1980} in the powerlaw spectrum. 
The problem of photon propagation through a moving medium has been studied
extensively in several works \citep{BlandfordPayne1981,Nobilietal1993,Titarchuketal2003}.
There are numerous studies on this context in the literature, both in theory and radiation hydrodynamic simulation \citep{Eggumetal1988,Kingetal2001,Okuda2002,Ebisawaetal2003,Ohsugaetal2005}.
In addition to down-scattering of hard radiation, jet can also behave like an up-scattering Comptonizing medium of the
soft photons from the disk. The presence of soft-excess is a commonly observed spectral component for both
X-ray binaries and AGNs. This can be originated if there is some warm Compton up-scattering
medium is present is the system. The excess component can also be associated as an 
intrinsic component with the disk, powered by the mass accretion rate \citep{Doneetal2012}, appeared
as a reflection component \citep[][for a review]{Doneetal2007}. 
This distortion of the power-law continuum above $\sim 10$ keV energy could also be due to down-scattering 
of hard radiation 
by the outflowing plasma rather than from a static reflecting medium (TS05). In addition to these spectral features, 
a converging flow falling onto the black hole near the event horizon can show bulk motion effect 
\citep[][hereafter CT95]{ChakrabartiTitarchuk1995,TitarchukZannias1998}.
Under such a complex radiation components coming from the accretion-ejection system,
it is important to ask, independent of how a jet is launched from the disk, what would be the overall 
X-ray spectrum in the presence of a jet, and what is the 
contribution from the base of the jet itself, specially the subsonic region which is hot. 
Most importantly, whether there is any evidence of such a contribution. 

\citet[][hereafter C99]{Chakrabarti99} postulated that the jets in the black hole systems are formed out of the 
post-shock region, which is also called the CENtrifugal pressure dominated BOundary Layer or CENBOL. This 
is the subsonic, hot region between the shock and the inner sonic point \citep{Chakrabarti1989} located just inside the marginally stable orbit.
C99 also proposed that the entire outflow is produced 
by the CENBOL and the mass outflow rate directly depends on the compression ratio $R = \frac {\rho _{+}}{\rho_{-}}$ of 
the shock itself, where $\rho_{-}$ and $\rho_{+}$ are the densities of the flow in the pre-shock and post-shock regions. 
It was shown that the ratio of the mass outflow rate to inflow rate is indeed a function of $R$ varies in a non-linear way apart from a geometric factor coming from the solid angles of the outflow and the inflow.
For a very strong shock, which forms farther out as in a hard state, the thermal driving force is low 
and the jet is not powerful. Similarly, in a soft state, when the shock is the weakest, the size of the base of the jet is negligibly small, thermal driving force is also low and hence the ratio is also low. Only in between the two states, the ratio is significant. 
Several hydrodynamic simulations showed that a few percent of accreting matter is indeed present in the jet \citep{Moltenietal1994,Chattopadhyayetal2004,Garainetal2012}.
\citet{SinghChakrabarti2011} studied jet properties of the flow when the energy dissipation is present at the shock and found lower efficiency of the jet formation when viscosity is high.

In \citet{Chakrabarti1997} two component flow scenario, it is established that the emission from the CENBOL decreases as the accretion rate of the Keplerian disk is increased, both because its size and temperature are reduced. 
This reduction of thermal pressure of CENBOL reduces the mass outflow rate. Thus, accretion rates, spectral states, and jet are interlinked.
Observations evidenced prominent jets in hard and hard-intermediate spectral states (HS and HIMS, \cite{Fenderetal2004}). 
Transonic solution including real cooling \citep{MondalChakrabart2013} and outflows \citep{Mondaletal2014b}
showed that spectral states are related to mass outflow rate and a significant change in energy spectral index 
are observed in HS and HIMS. These solutions are self-consistent in the sense that cooling and mass loss are 
incorporated in the flow equations before obtaining the spectra. Recent work by \citet{NagarkotiChakrabarti2016} using a viscous solution confirms 
that a significant amount of mass loss is centrifugally driven as suggested by \citet{BlandfordPayne1982} though the disk may be sub-Keplerian in nature. 

The interdependency between the accretion rate, CENBOL size and QPOs is also well known in the literature 
During an outburst, when a source moves from HIMS to SIMS, the QPO type switches from
type-C to type-B \citep[e.g.,][and references therein]{Mondaletal2014,Debnathetal2015,Chatterjeetal2016}. 
However, jet properties also changed drastically, which is observed in large amplitude radio flares \citep{Fenderetal2004,Fenderatal2009}. \citet{RadhikaNandi2014} reported the accretion-ejection mechanism for the object XTE~J1859+226 and 
pointed out that the quasi-periodic-oscillations (QPOs) disappear during the high ejection of jets. All these are expected from the {\fontfamily{qcr}\selectfont TCAF} paradigm, since QPOs are
known to be formed due to the resonance of various time scales inside CENBOL which 
also produces jets. Interestingly, the highest outflow to inflow
rate from the theoretical solution of C99
happens for intermediate compression ratio of the shock, i.e., in hard and soft-intermediate states. Retaining CENBOL is possible only in a low viscosity state \citep{Mondaletal2017}.

The original {\fontfamily{qcr}\selectfont TCAF} solution of CT95 and C97 does not include the contribution from outflows
and therefore, presence of significant outflows require a modification of the
model. This is the primary goal of this paper. 
Here, we concentrate on the effects on the emergent
spectra when the base of the outflow also participates in Comptonization of injected photons along with the CENBOL. 
We study the jet spectrum when the accretion rates, the geometry of the 
flow and jet parameters vary. We also see the variation of the spectra from the jet with the variation 
of CENBOL size and shock compression ratio. The study of optical and radio observations \citep{Brocksoppetal2001,Paharietal2017} inferred the presence of jet in GS\,1354-64 system, however, there is very little study on this object as far as the effects of jets are concerned. We apply the solution to study this
object, in order to throw some light on accretion-ejection behavior during its 2015 outburst. In fact, from the fits of the modified {\fontfamily{qcr}\selectfont TCAF} model, we extract the
jet parameters, such as the mass outflow rate, as well. In future, we aim to 
apply the solution to other well known candidates which exhibit prominent jets.

The paper is organized as follows: in the next Section, we present the configuration of the jet and the equations we used to calculate
jet temperature and internal number density, etc. In \S \ref{sec:Results}, we present the results of the effects of mass loss on normal in {\fontfamily{qcr}\selectfont TCAF} output,
mainly the spectral variation with location of the shock, shock compression ratio and accretion rates. In \S \ref{sec:Gs1354}, we calculate the mass 
outflow rate of the BHC GS~1354-64 during its 2015 outburst using {\it NuSTAR} observation. Finally, in \S \ref{sec:Conclusion}, we briefly present our concluding remarks.

\section {Equations and jet configuration}\label{sec:JetConfigEqn}
\begin{figure}
\centering{
\includegraphics[height=6.0truecm,angle=0]{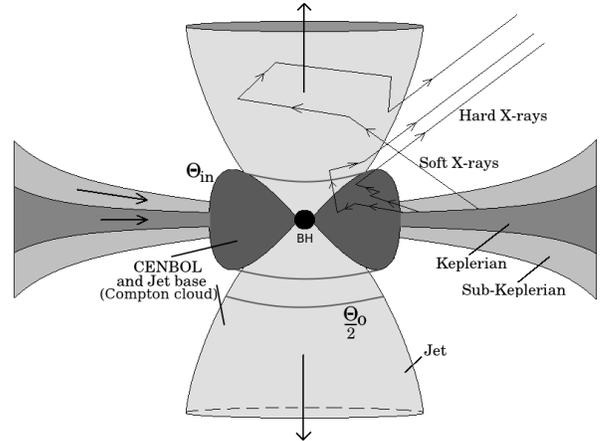}}
\caption{A cartoon diagram of the geometry of an accretion flow with a jet in the vertical direction. Here, $\Theta_{\rm o}$ and $\Theta_{\rm in}$ are solid angles subtended by the outflow
and the CENBOL respectively. The base of the jet is treated as a Comptonizing medium, which also contributes in computing model spectra. Zigzag trajectories are the typical paths followed by soft photons from the Keplerian disk. These photons are Comptonized by the CENBOL and the jet base (marked) and are radiated as hard X-rays.
} 
\label{fig:cartoon}
\end{figure}

To extract the jet properties from a spectral fit, we consider a conical jet in the vertical direction of the flow, which is originated from the post-shock region. 
The base of the jet is a part of the Compton cloud along with the CENBOL.
In \autoref{fig:cartoon}, we present a cartoon diagram of the two-component flow with a jet where a Keplerian disk resides at the equatorial 
plane, and it is surrounded by the low angular momentum flow (dynamic halo). At the center, a black hole of mass $M_{\rm BH}$ is located. We show also the solid angles $\Theta_{\rm o}$ and $\Theta_{\rm in}$ subtended by the two outflow components and the axisymmetric torus created by the CENBOL respectively.

The spectrum emitted from the system contains (a) modified blackbody component, which is coming from the Keplerian disk, (b) 
the soft photons from the disk are up-scattered by the hot electrons at the inner region of the disk and come out as hard radiation, 
(c) disk soft photons also get up-scattered by the jet base, which modifies the spectrum on the shoulder of the blackbody bump, and (d)
hard radiation from the CENBOL are down-scattered by the outflowing jet due to Doppler effect. The components (a) and (b) are already computed in the original
\citet{ChakrabartiTitarchuk1995} paper. For the component (c) we use the same radiative transfer calculation incorporating the 
temperature and optical depth of the 
jet. Once we have the
 physical quantities of the jet medium and the spectral component (b), we use it to 
pass through the jet medium following TS05. 
Below, we discuss the equations used
to estimate different physical quantities of the jet medium.

The radiation from the CENBOL may constantly heat up the jet and the jet 
may not get enough time to cool at least up to the sonic point. Thus, the ratio of the mass outflow rate to 
the mass inflow rate is derived for an isothermal base of the jet (C99), which is given by,
\begin{equation}
R_{\dot m}=f_{\rm col}f_0^{3/2}\frac{R}{4} exp\left[\frac{3}{2} -f_0\right],
\end{equation}
where, $f_{\rm 0}=\frac{R^2}{R-1}$ and $f_{\rm col}$(=$\Theta_{\rm o}/\Theta_{\rm in}$) is the 
collimation factor. Here, $\Theta_{\rm o}$ and $\Theta_{\rm in}$ are the solid angles subtended by the two outflow components and the inflow.
We should mention that $f_0$ factor diverges when R$\sim$1. However, that does not
create any complications in producing the spectrum, since when 
 R$\sim$1, the flow is not a {\fontfamily{qcr}\selectfont TCAF} and produces the blackbody spectrum only. 
It is clear that the mass outflow rate is a function of only $R$ and $f_{\rm col}$. Here, $f_{\rm col}$ is a parameter 
and its value is varied in the range (see \autoref{table:lmodel}) to fit the observed data. One of the values of $f_{\rm col}$ is $0.1$, 
which corresponds to $\Theta_{\rm o} \sim 10^{\circ}$.  
\citet{Moltenietal1994} discussed that $\Theta_{\rm o}$, should depend on the strength of the centrifugal barrier.
If the barrier is stronger, $\Theta_{\rm o}$ and consequently the mass loss rate is higher.
The total accretion rate of the in-flowing matter at the CENBOL after mixing of the Keplerian and sub-Keplerian components is given by,
\begin{equation}
\dot M_{\rm in}=\dot M_{\rm d}+R \dot M_{\rm h},
\end{equation}
here, $\dot M_{\rm d}$ and $\dot M_{\rm h}$ are the disk and halo rates respectively of the flow in gm\,sec$^{-1}$ unit.
Since in {\fontfamily{qcr}\selectfont TCAF}, the boundary layer, i.e., CENBOL, is emitting the jet, the initial launching velocity of the ejecta comes
from the temperature of the post-shock region, i.e., ${\rm v}_{\rm j}\sim \sqrt T_{\rm shk}$ $\sim 0.1c$. However, it changes depending 
on the CENBOL location. 
The density at the base of the jet is calculated using,
\begin{equation}
\rho_{\rm j0}=\frac{\dot M_{\rm j}}{\pi X_{\rm s}^2 {\rm v}_{\rm j}},
\end{equation}
where, $\dot M_{\rm j}(=R_{\dot m}\dot M_{\rm in})$ is the rate of the 
mass outflow. As the height increases, density of the outflowing matter falls as, 
\begin{equation}
\rho_{\rm j}=\rho_{\rm j0}\left(\frac{r_{\rm j}}{r_{\rm jb0}}\right)^{-2},
\end{equation}
where, $r_{\rm j}$ and $r_{\rm jb0}$(=h$_{\rm shk}$) are the (running) height of the jet and 
the fixed height of the base of the jet respectively. The latter is 
basically the height of the shock (h$_{\rm shk}$). We assume that the jet matter moves away 
subsonically till the sonic surface, the radius of which is given by (C99),
\begin{equation}
r_{\rm c}=\frac{f_{\rm 0} X_{\rm s}}{2}.
\end{equation}
We assume that the temperature of the jet ($T_{\rm j}$) varies as $\sim$ $1/r_{\rm j}$. In this work, 
we also consider the effects of down-scattering of the Comptonized photons by the bulk motion of the 
outflowing jet as discussed in TS05. There are a number of sources which show 
evidence of strong reflection above $\sim 10$~keV.
This distortion of the power-law continuum above this energy can also be originated from the down-scattering of the hard radiation by the 
outflowing plasma rather than from a static reflecting medium. This is discussed in CT95 and TS05. 
Here, we implement this to explain the excess radiation component present in the spectrum. 
We use the following equation from TS05 for the final down-scattered jet spectrum:
\begin{equation}
F_{\rm BMC}\propto \Phi_{\rm Comp}\lbrace 1 + N_{av}z[(1-\alpha)-\frac{\epsilon}{3z_\ast}-\frac{z}{z_\ast}] \rbrace,
\end{equation}
where the second term in the bracket describes the pile-up and softening of the input Comptonization spectrum ($\Phi_{\rm Comp}$), resulting 
from the down scattering effect by the jet electrons.
Here, $N_{\rm av} (= 3 \tau_0^2 /8)$ is the number of average scattering, 
$\epsilon$ is the efficiency of the energy loss
in the divergent flow $(\sim 2{\rm v}_{\rm j}/c\tau_0)$, $z$ is the dimensionless photons energy, $z_\ast$ is the dimensionless temperature and $\tau_0$ is the optical depth 
of the scattering medium, namely, the base of  the jet. 
The $\Phi_{\rm Comp}$, we used here, is
the spectrum emitted from the CENBOL due to up-scattering of disk soft photons and
$\alpha$ is the energy spectral index estimated from the original {\fontfamily{qcr}\selectfont TCAF} code (CT95) to make the computation to be consistent. 
However, TS05 assumed this spectral component to be  
a cut-off power-law for a given $\alpha$. The subscript BMC implies the bulk motion Comptonization.

\begin{table} 
\scriptsize
\centering
\caption{\label{table:lmodel} {\fontfamily{qcr}\selectfont JeTCAF} model parameters range}
\begin{tabular}{lccccccr}
\hline
Pars.     &Units        &Default &Min.&Min. &Max.&Max. &Increment   \\
\hline
M$_{\rm BH}$          & M$_\odot$       &6.0   &4.0   &4.0   &15.0  &15.0  &2.0   \\
$\dot m_{\rm d}$      & $\dot M_{\rm Edd}$  &0.01  &0.001 &0.001 &2.0   &2.0   &0.05   \\
$\dot m_{\rm h}$      & $\dot M_{\rm Edd}$  &0.1   &0.001 &0.001 &5.0   &5.0   &0.2   \\
$X_{\rm s}$           & $r_{\rm g}$           &50.0  &6.0   &6.0   &400.0 &400.0 &8.0   \\
R                     & `` ''           &2.1   &1.1   &1.1   &7.0   &7.1   &0.2   \\
f$_{\rm col}$         & `` ''           &0.1   &0.05  &0.05  &0.5   &0.5   &0.01   \\
\hline
\end{tabular}
\end{table}

The units of mass, length, speed, and accretion rates are 
$M_\odot$, $r_g=2GM/c^2$, $c$, and $\dot M_{\rm Edd}$ respectively, 
where $M_\odot$, $G$, $M_{\rm BH}$, $c$ and $\dot M_{\rm Edd}$ are the mass of the Sun, gravitational constant, mass of the central object, 
speed of light and Eddington accretion rate. Throughout the paper we follow these units for model parameters and physical quantities.

\section{Results and Discussions} \label{sec:Results}
\begin{figure*}
\hspace{-0.9cm}
\centering{ 
\includegraphics[height=7.truecm]{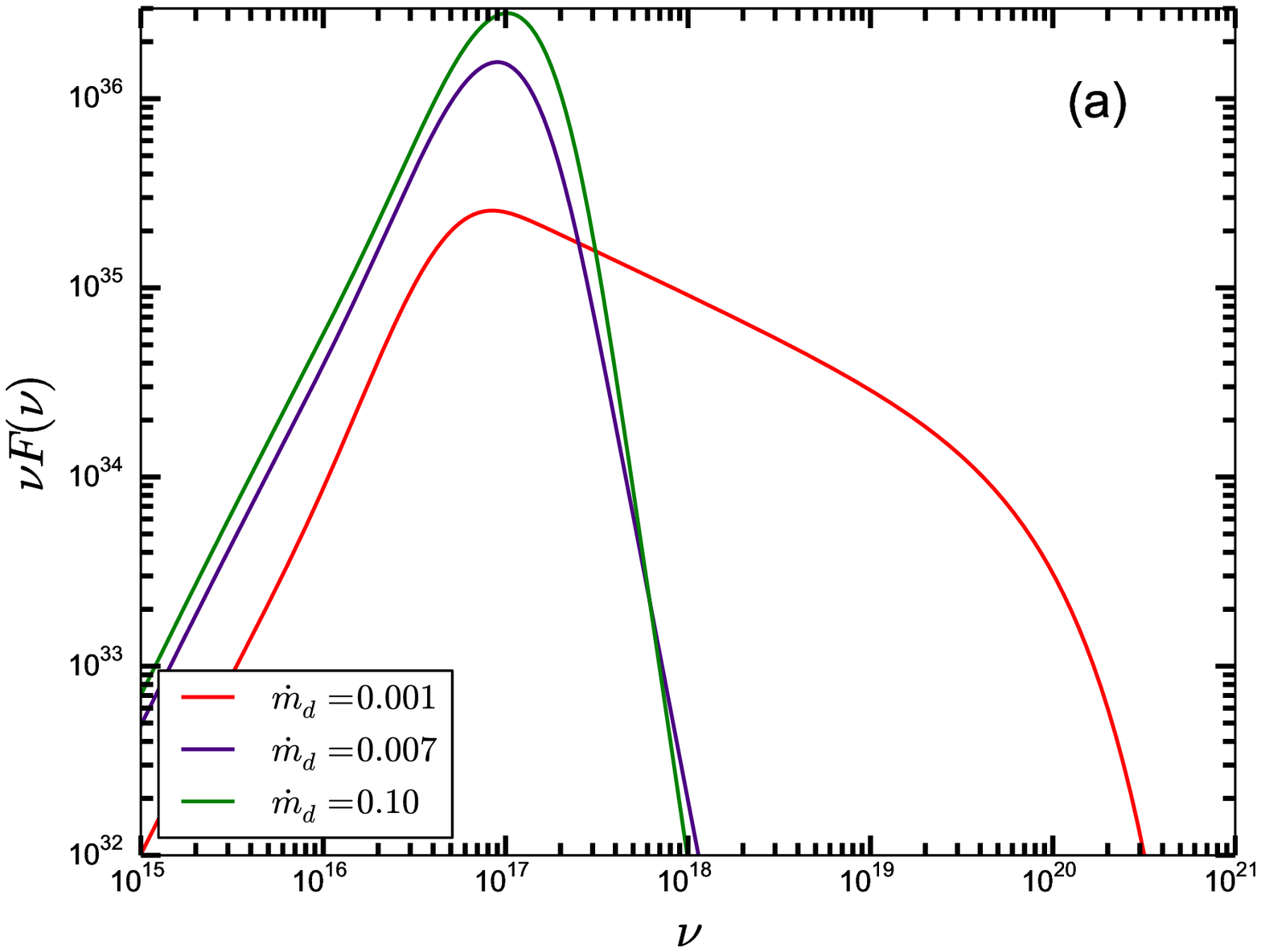}
\hspace{-0.9cm}
\includegraphics[height=7.truecm]{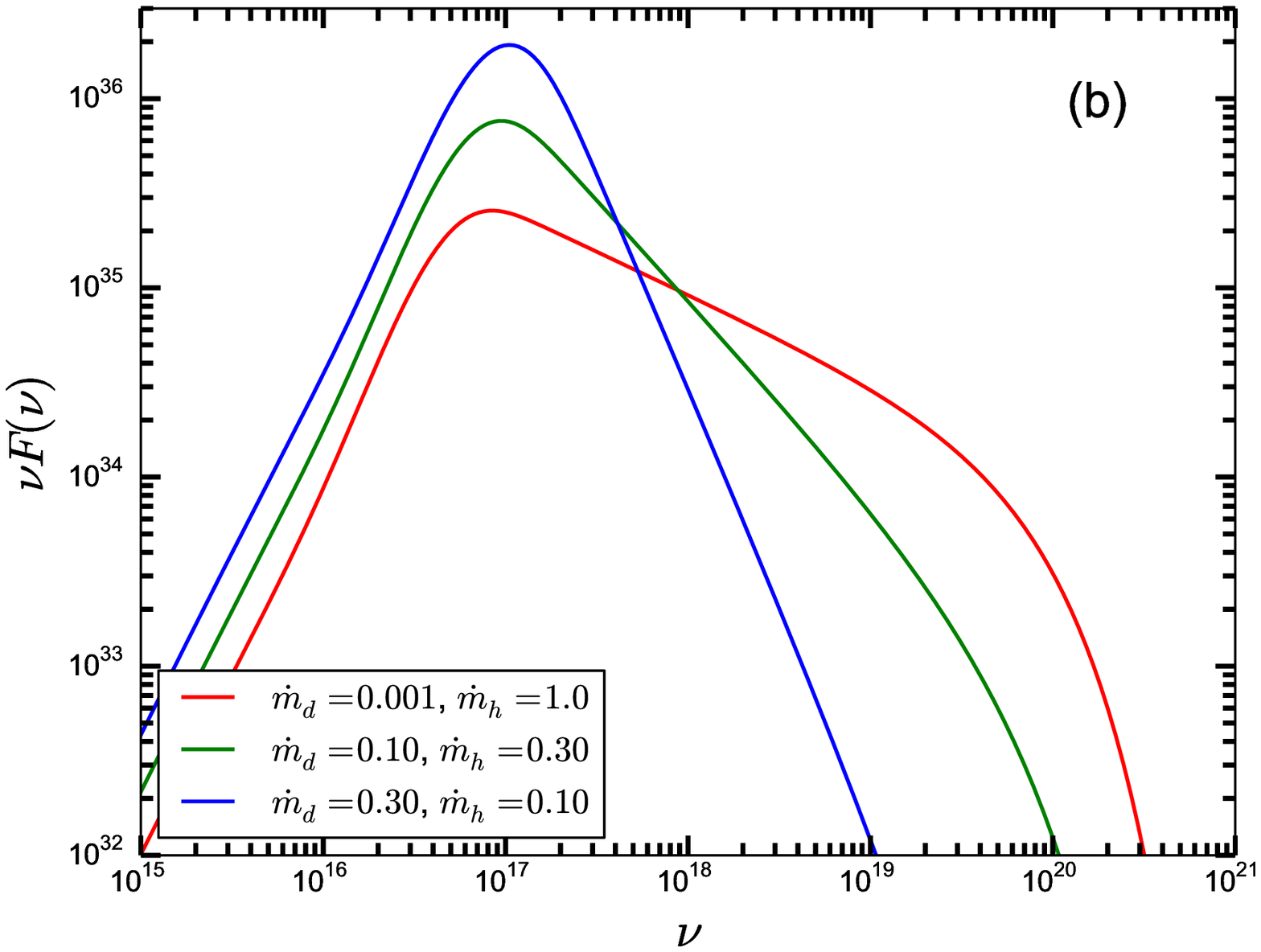}}
\caption {(a) Scattering of soft photons from the disk by the jet when disk accretion rate varies from 0.001 to 0.1 when halo rate is 1.0, keeping all other flow parameters fixed. The spectrum softens with increasing accretion rate. (b) Similar spectral variation as in (a) for 
different sets of disk and halo accretion rates, when other parameters are the same as in (a). The spectrum softens when the accretion
rate changes to change the spectral state from hard to intermediate. 
Other parameters are $M_{\rm BH}$=10.0, $X_{\rm s}$=20.0, 
R=2.5 and $f_{\rm col}$=0.1. Here, frequency is in Hz unit.}
\label{fig:VaryMdHsSs}
\end{figure*}

We show how the spectral properties change when the base of the jet is also a part of the Compton cloud. 
In \autoref{fig:VaryMdHsSs}a, we show the spectral variation of the jet component from the total emitted spectrum
for different accretion rates ($\dot m_{\rm d}$), when other flow
parameters ($\dot m_{\rm h}$, $M_{\rm BH}$, $X_{\rm s}$ and $R$) are kept constant. We increase the accretion rates, i.e., the number of soft photons and observe that the jet spectrum softens. Here, we used $f_{\rm col}$=0.1 and 
other parameters are marked with figure. Increase in soft photon number also cools down the Compton cloud rapidly. 
Thus, the emitted jet does not have sufficient 
thermal energy to up-scatter the soft photons significantly.
In \autoref{fig:VaryMdHsSs}b, we present the spectra of different spectral states 
by varying the accretion rates only, 
keeping all other parameters fixed. Here the spectra
vary from HS to SS and the same variation is reflected in the jet contribution in the spectra as well. 
Spectral states may change due to different sizes of the cloud and compression ratio of the shock, since, these two physical parameters are also related to density and optical depth of the system.

\begin{figure*}  
\hspace{-1.0cm}
\centering{
\includegraphics[height=5.0truecm]{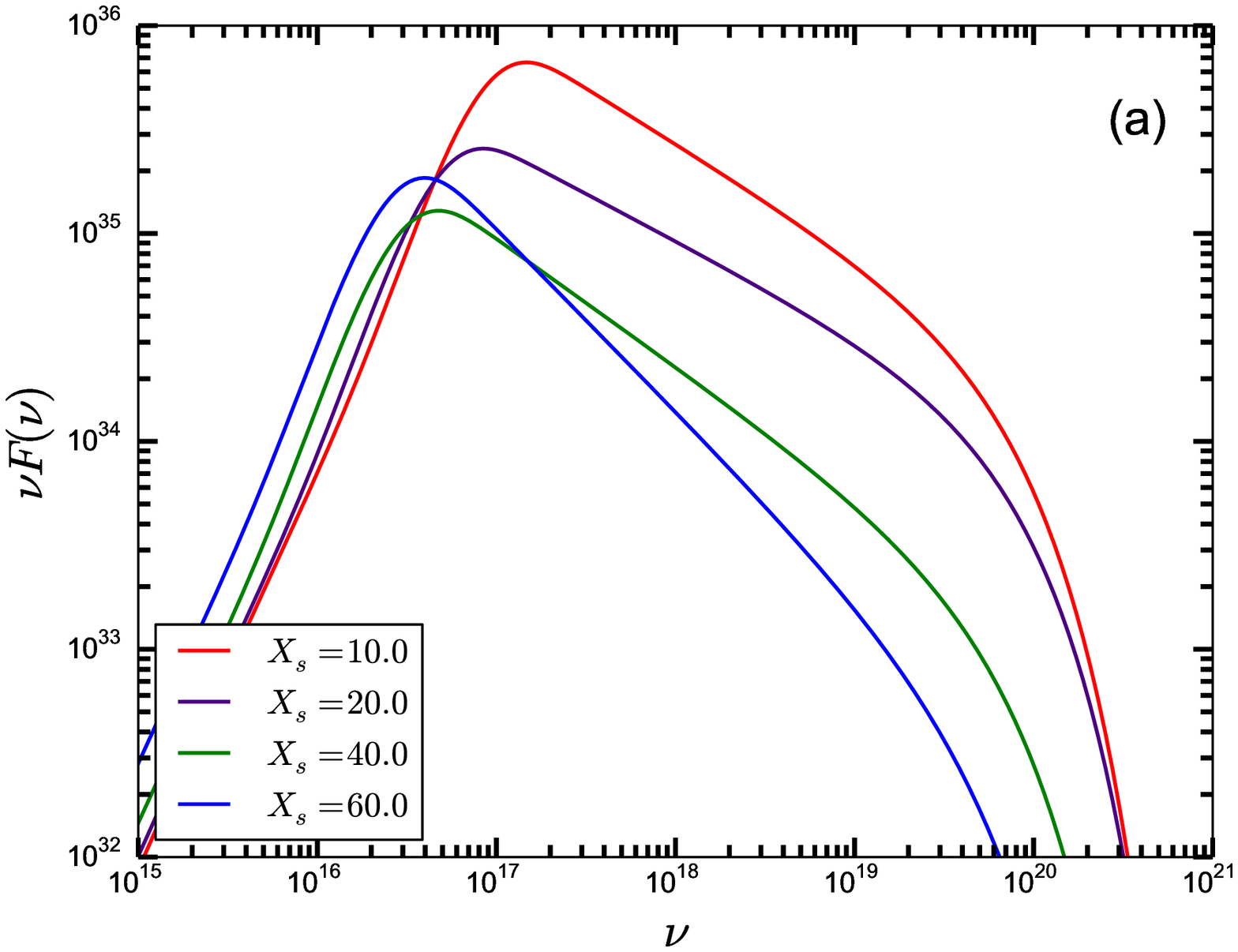}
\hspace{-0.8cm}
\includegraphics[height=5.0truecm]{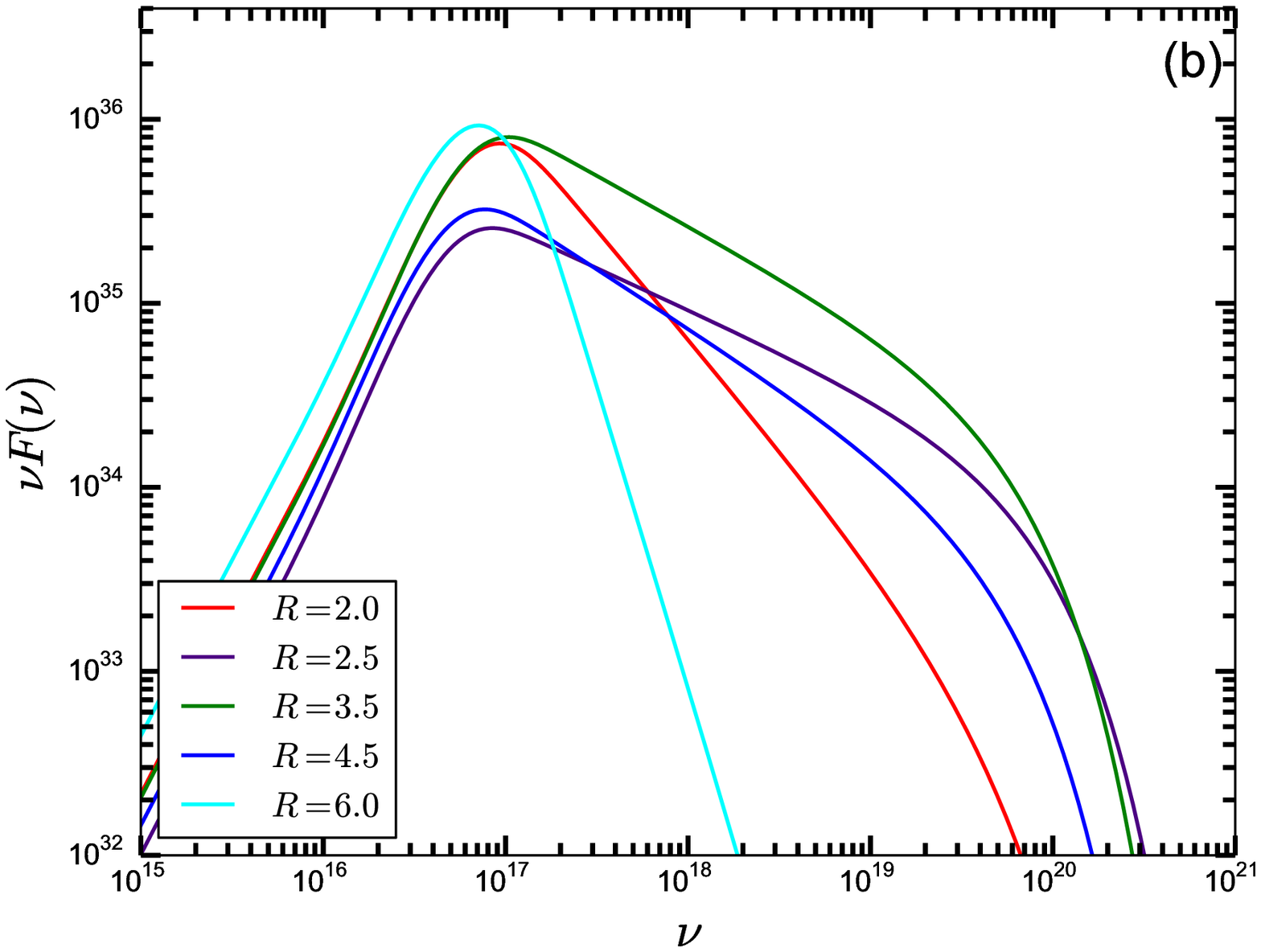}
\hspace{-0.8cm}
\includegraphics[height=5.0truecm]{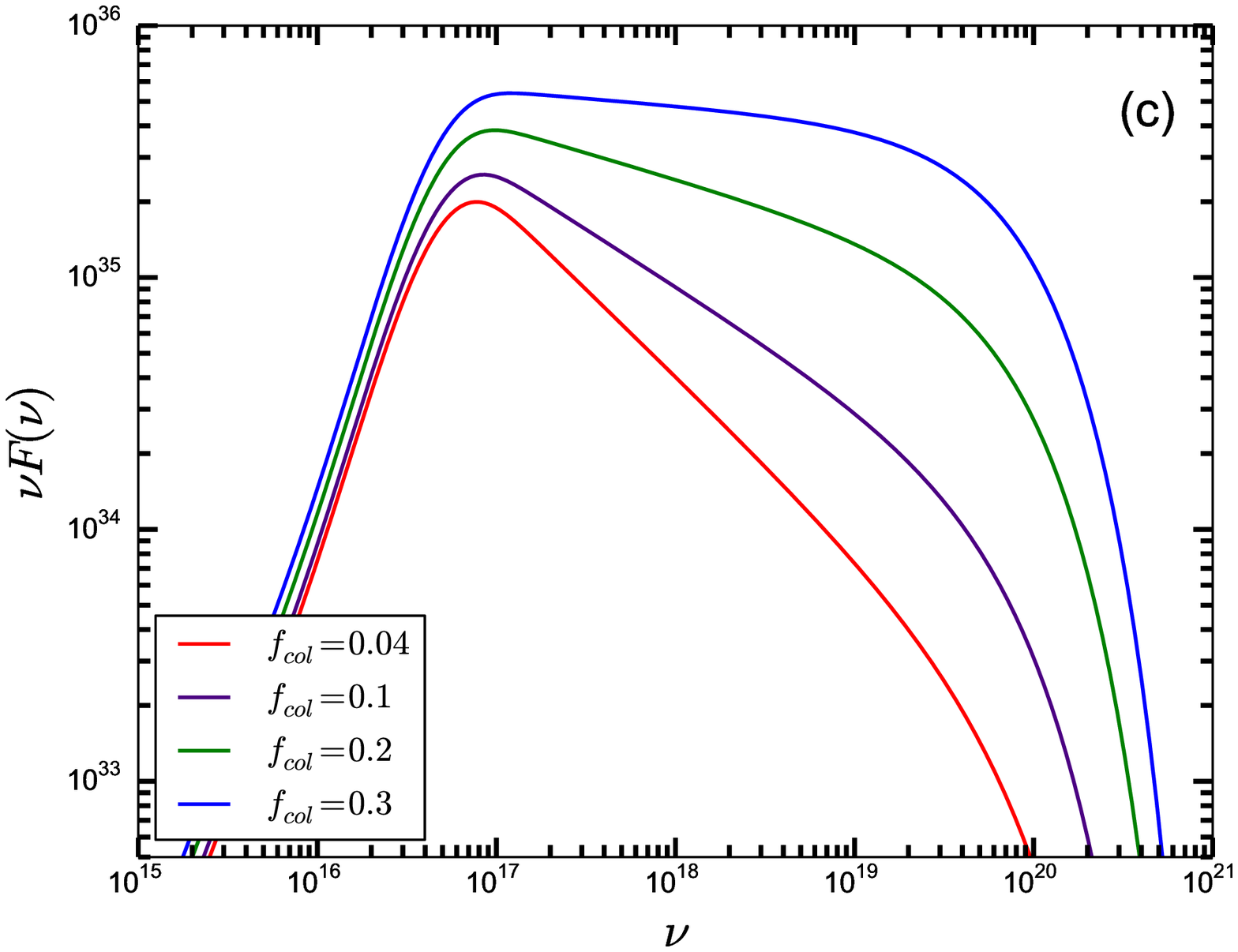}}
\caption{Same as \autoref{fig:VaryMdHsSs}, (a) for varying size of the Compton cloud ($X_s$) when R=2.5 and $f_{\rm col}$=0.1, (b) for different values of 
compression ratio R when $X_s$=20.0 and $f_{\rm col}$=0.1, and (c) for different values of collimation factor ($f_{\rm col}$) when $X_s$=20.0 and R=2.5. In (a), as the 
location/size increases, the spectrum becomes softer. In (b) as the value of R increases, mass outflow rate also increases,
which increases the optical depth and the average number of the scattering. Thus, the spectrum 
becomes harder. However, mass outflow rate starts to decrease when R becomes higher than 4.5 and the spectrum again gets softer.
This can be seen for $R=6.0$. In (c), spectrum becomes harder with increasing collimation factor.
The other parameters for both the plots are $M_{\rm BH}$=10.0, $\dot m_{\rm d}$=0.001, and $\dot m_{\rm h}$=1.0.
} 
\label{fig:VaryxsRfcol}
\end{figure*}

\autoref{fig:VaryxsRfcol}a shows the effects of the size of the Compton cloud ($X_{\rm s}=10,\ 20,\ 40$ and $60$) on the jet component of the total emitted spectrum. 
We see that as the size of the cloud increases, the jet spectral component becomes softer. 
As the base of the jet is the CENBOL, the jet component varies with shock location.
Larger CENBOL being farther from the black hole, its temperature is already less (roughly $\propto 1/X_s$). Also, it intercepts more soft photons from the Keplerian disk which reduces the overall CENBOL temperature \citep{Chakrabarti1997}. Thus, the resulting temperature of the base of the jet becomes lower. 
Thus, the soft primary photons are not up-scattered many times. This causes the jet spectrum to be softer. 
In \autoref{fig:VaryxsRfcol}b, we show the jet component variation for different values of $R$ (=$2.0$, $2.5$, $3.5$, $4.5$ and $6.0$). We 
observe significant changes in the jet component, which becomes softer with increasing R. From C99 solution, we see that
the mass outflow rate decreases when $R$ is higher than $\sim 4.0$. This affects the spectral shape, for instance, when $R=6.0$, 
mass outflow rate is very low and consequently the spectrum is also softer. In \autoref{fig:VaryxsRfcol}c,
we see the spectral hardening when jet collimation factor ($f_{\rm col}$) has changed from $0.05$ to $0.3$.  

\begin{figure*}
\hspace{-0.7cm}
\centering{
\includegraphics[height=7.truecm]{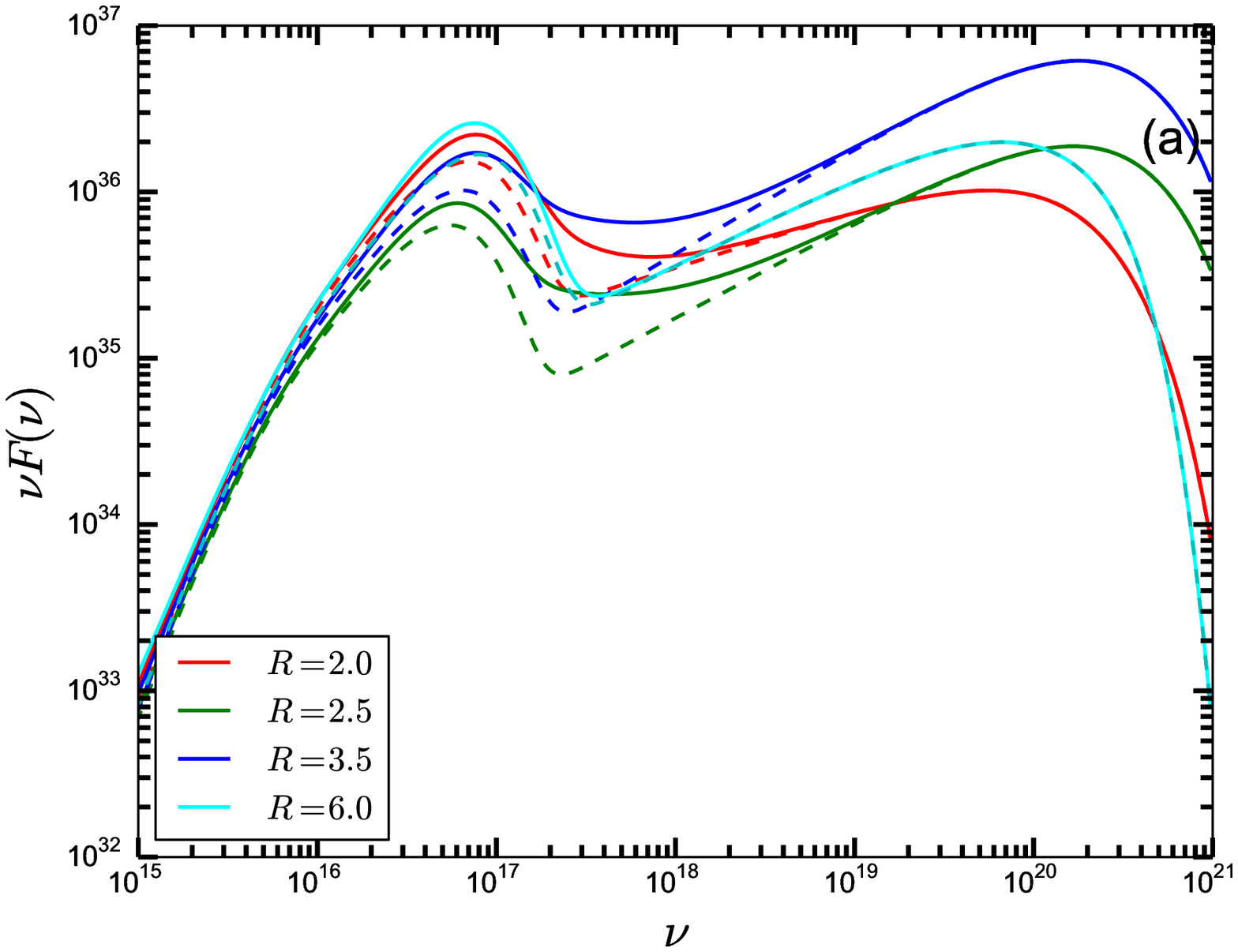}
\hspace{-0.9cm}
\includegraphics[height=7.truecm]{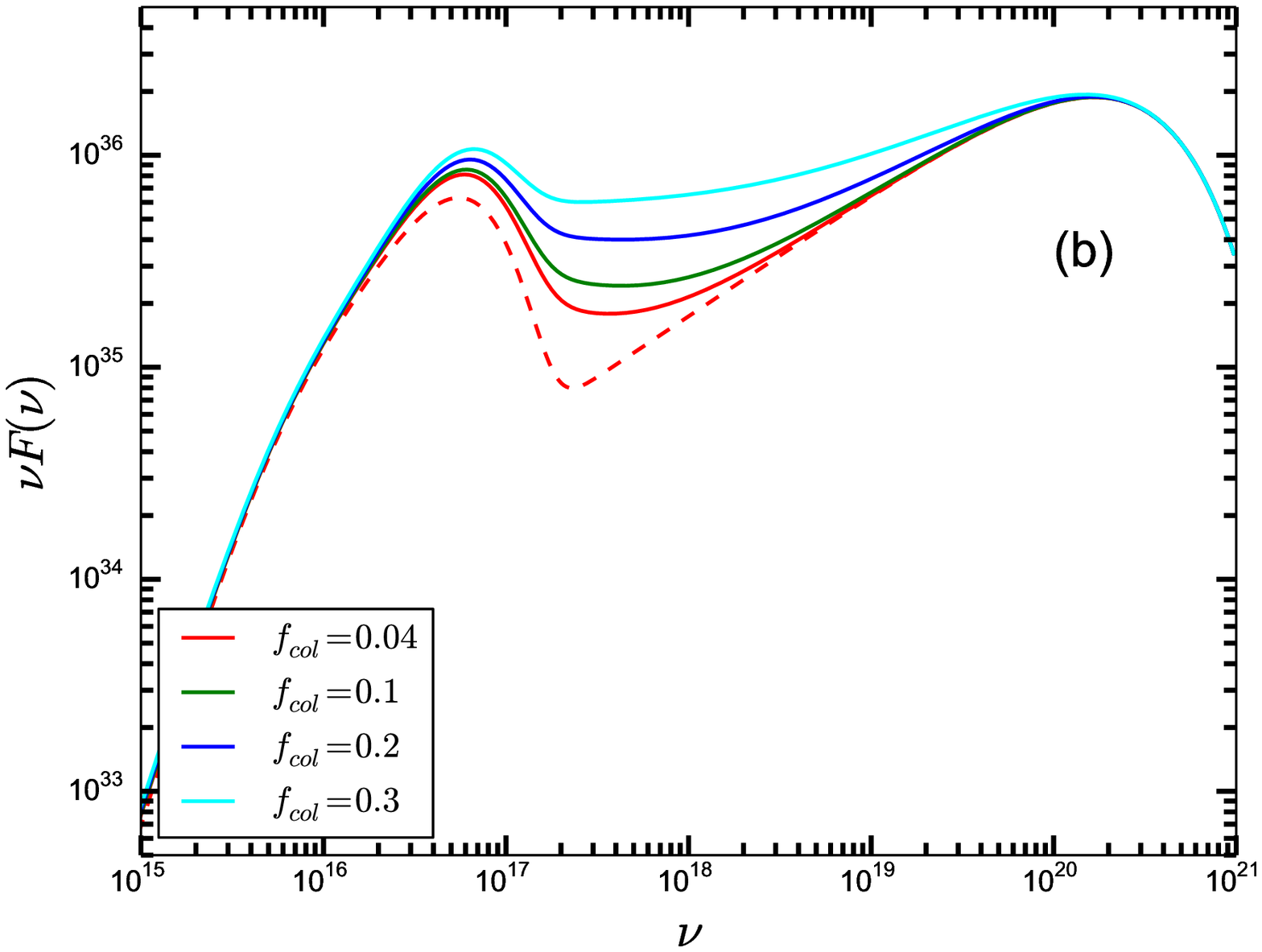}}
\caption{(a) Total emitted spectra with (solid lines) and without the jet contribution (dashed lines, same as {\fontfamily{qcr}\selectfont TCAF} spectra), 
	for different values of $R$, and (b) $f_{\rm col}$. In (a), we see that 
	the spectrum becomes harder with increasing $R$.
	The new parameter ($f_{\rm col}$) does not affect the CENBOL
	spectrum and thus the total spectrum without the jet contribution is the same for all (dashed line).
	The other parameters are $M_{\rm BH}$=10.0~$M_\odot$, $\dot m_{\rm d}=0.001$, $\dot m_{\rm h}=1.0$ and $X_{\rm s}=20.0$. Here, frequency is in Hz unit.
}
\label{fig:TotalRfcol}
\end{figure*}

\autoref{fig:TotalRfcol}a shows the variation of the net spectrum with (solid) and without jet (dashed line) contribution as a function of the
shock compression ratio ($R$). We note that due to the presence of the jet, the total spectrum becomes harder 
as compared to that without the jet. The variation of the total spectrum with $R$ is significant. In \autoref{fig:TotalRfcol}b, we show the total 
emitted spectra for different values of collimation factor. The dashed line shows the total spectrum when jet contribution
is not included, and the solid colored lines are the total spectrum with the jet contribution. One can see that as the collimation factor
increases, the spectrum becomes harder when the other parameters are fixed. 

\begin{figure*}
\hspace{-0.7cm}
\centering{
\includegraphics[height=7.0truecm,angle=0]{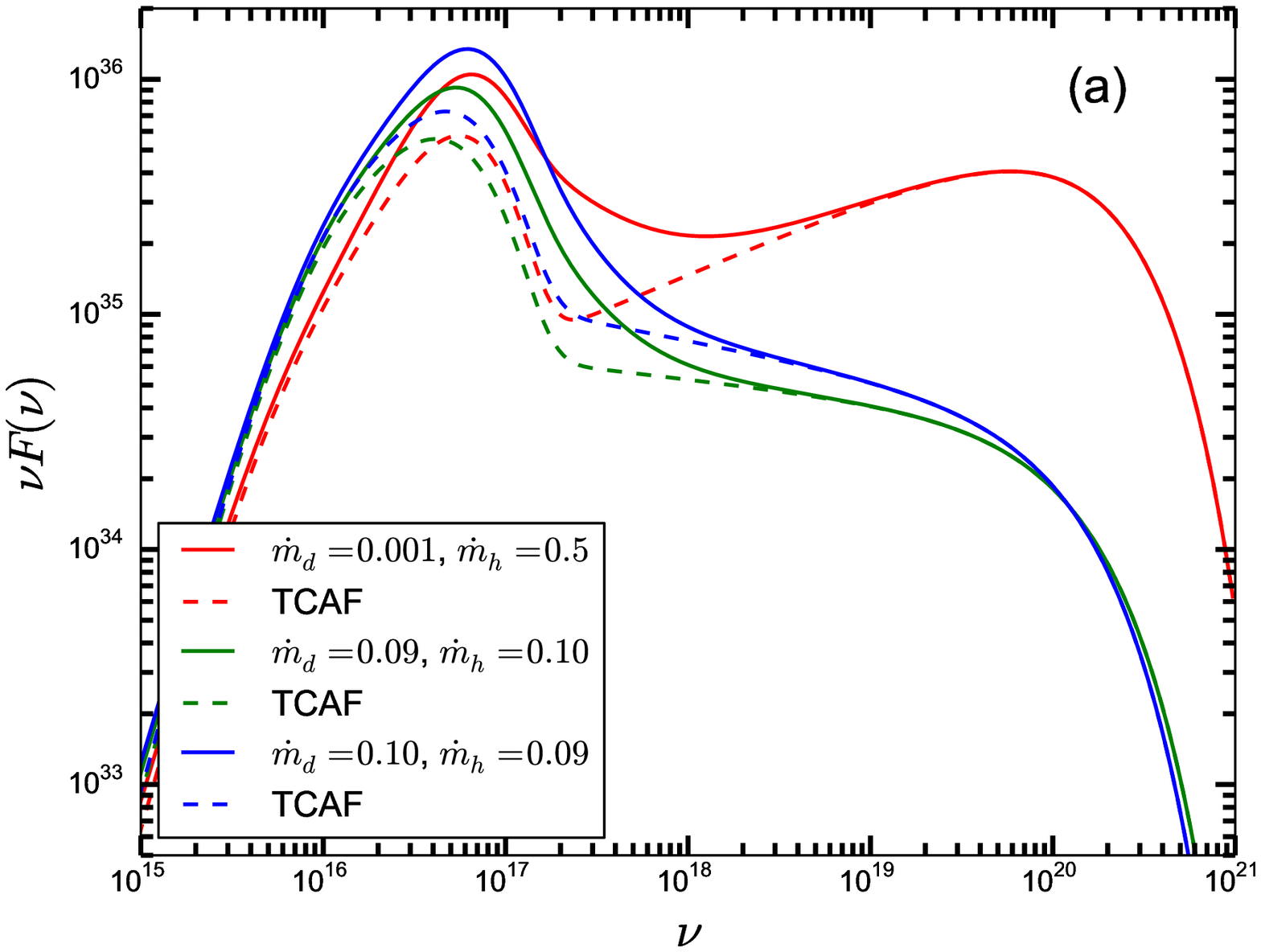}
\hspace{-0.9cm}
\includegraphics[height=7.0truecm,angle=0]{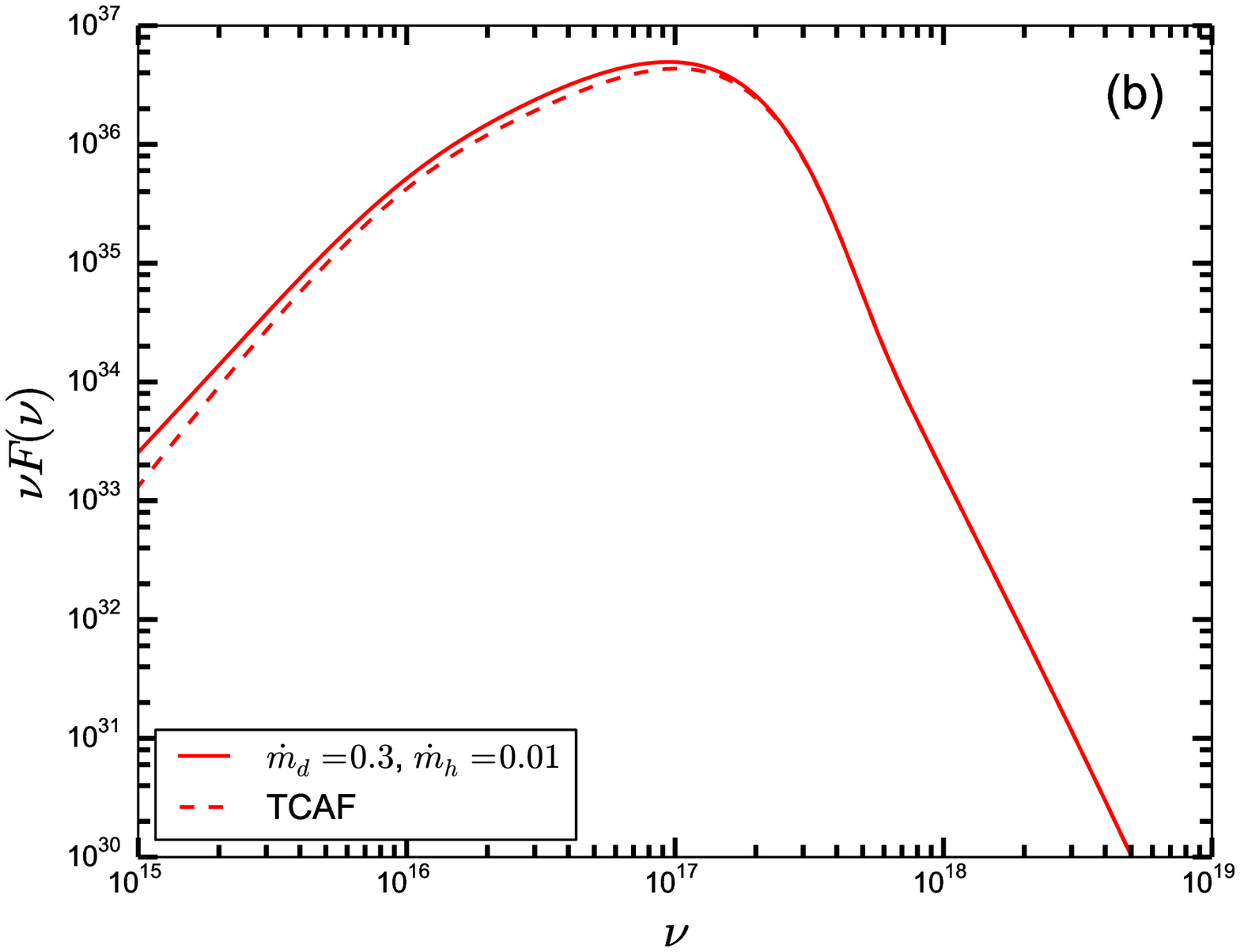}}
\caption{(a) Total emitted spectrum with (solid lines) and without (dashed lines) jet contribution for different spectral states
	depending on the accretion rates when other parameters are fixed ($X_{\rm s}=20.0$ and $R=3.0$). 
There is a significant change in the spectrum as 
	compared with the soft state in presence of jet. (b) Total spectrum in the soft state in presence of jet (solid line) 
component when $X_{\rm s}=10.0$ and $R=1.2$. 
	In the soft state, the jet contribution is negligible and both the spectra are the same.
	For both the plots $M_{\rm BH}$=10.0 and $f_{\rm col}=0.2$.  Here, frequency is in Hz unit.
} 
\label{fig:TotalStatesSoft}
\end{figure*}

\autoref{fig:TotalStatesSoft}a shows the total spectral variation of different states with (solid lines) and without (dotted lines) the jet contribution. 
Here the spectral state changes with accretion rates when other parameters of the model are kept fixed. 
As the accretion rate increases, more soft photons from the Keplerian disk are intercepted by the Compton cloud 
and the spectrum becomes softer. On the other hand, an increase in disk accretion rate cools down the CENBOL which reduces the thermal pressure
and thus reduces the mass outflow rate. Consequently, the system moves to the soft state through intermediate states. 
In \citet{Chakrabarti1996}, it was proposed that viscosity is mainly 
responsible for changing the relative rates, which is indirectly taken
into account by two accretion rates in {\fontfamily{qcr}\selectfont TCAF}. 
In \citet{Mondaletal2017}, it was shown that the observed viscosity parameter indeed
rises in the rising phase and decreases in the declining phase, thereby completing the full cycle of the
observed states.
In \autoref{fig:TotalStatesSoft}b, we choose typical model parameters for soft states. One can see that the jet contribution is negligible.
There are many observations in the literature which show the correlation between observed spectral states with the jet \citep{Fenderetal2004,Bellonietal2005,Srirametal2012}.
They also discussed correlations between the QPOs and the jet emissions in the 
HID diagram. All these correlations fall in place once the computation of the outflow 
rate to inflow rate ratio as a function of the properties of the centrifugal barrier was made (C99). The theoretical result directly shows that 
if indeed the outflows are produced from CENBOL, then,
the soft and extremely hard states are going to produce very little outflows. Only intermediate states
have higher outflow rates for a given inflow rate. When the optical depth is higher at the base of the jet, it produces blobby
jets \citet{Chakrabartietal2002,Nandietal2001}. In the soft state, due to cooling effects, this region is quenched and 
the mass outflow rate is reduced \citep[see also,][]{Garainetal2012}. Recently, \citet{Janaetal2017} estimated the X-ray flux 
from the base of the jet for Swift~J1753.5-0127 BHC using {\fontfamily{qcr}\selectfont TCAF} model.
 
\begin{figure}
\hspace{-1.0cm}
\centering{
\includegraphics[height=7.0truecm]{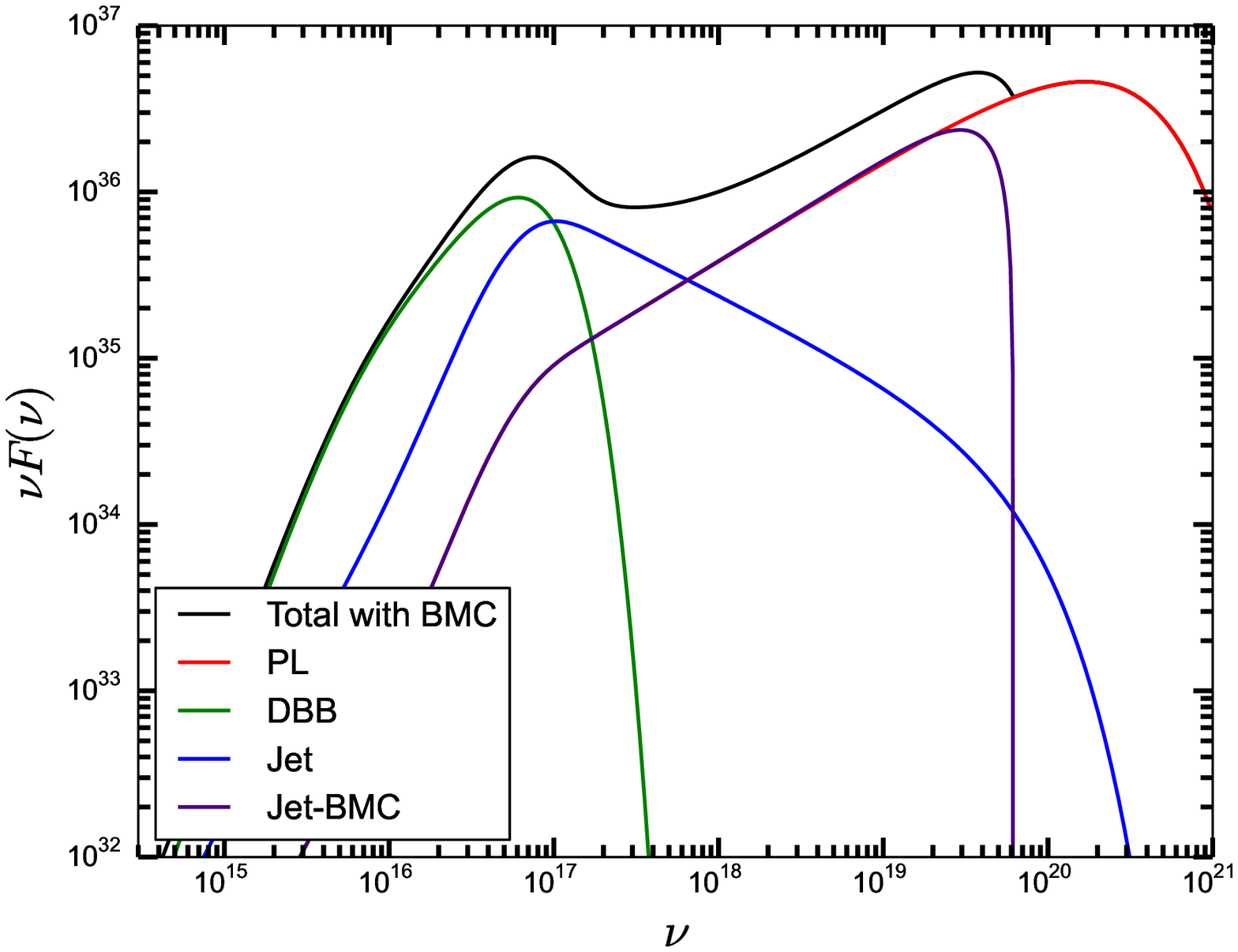}}
\caption{Different components including the jet which contribute to the net spectrum. Black line shows the total spectrum with a jet, the blue line 
is the jet component and the indigo line indicates the BMC effect of the outgoing jet.
Red and green lines show thermal Comptonization and blackbody spectra as in the original {\fontfamily{qcr}\selectfont TCAF} solution. 
The disk and the jet parameters for this plot are $M_{\rm BH}$=10.0, $\dot m_{\rm d}=0.001$, $\dot m_{\rm h}=1.0$, $X_{\rm s}=20.0$, 
$R=3.0$ and $f_{\rm col}=0.1$. Here, frequency is in Hz unit.
} 
\label{fig:Components}
\end{figure}

Till now, we did not show the contribution of BMC component in the spectral plots. In \autoref{fig:Components} we show the 
emitted spectrum when hard photons from the CENBOL pass through the outflowing diverging jet.
In \autoref{fig:Components}, we show the spectral components including jet  
when the down-scattering effects of the CENBOL photons due to the bulk motion of the jet is considered. 
The black line shows the net spectrum when the BMC effect is also included, the blue and indigo lines show 
the jet component with BMC effect respectively. Here, green and red lines are disk blackbody and thermal Comptonization spectra obtained from the original 
C97 solution. The net spectrum clearly shows a bump, which is slightly higher in {\fontfamily{qcr}\selectfont JeTCAF} spectra as compared to the {\fontfamily{qcr}\selectfont TCAF}. 
In, TS05 explained that the bump is due to the BMC effects of the CENBOL photons by the diverging outflows discussed in CT95.
In the next Section, we use both {\fontfamily{qcr}\selectfont TCAF} and {\fontfamily{qcr}\selectfont JeTCAF} model to fit the observed data, where we find the signatures of the BMC effect.

\section{GS~1354-64} \label{sec:Gs1354}
We now use our computed spectra to fit the data of the black hole candidate GS~1354-64.
GS~1354-64 is a dynamically confirmed low mass X-ray binary with a black hole mass $>$~7~$M_{\odot}$, 
$\sim 2.5$ day orbital period, and a distance of $\sim 25-61$~kpc 
\citep{Casaresetal2004,Casaresetal2009}. 
\citet{ReynoldsMiller2011} reported the possible  
distance to be $\sim 15$~kpc and this large difference is due to the extinction effects. 
Very recently, \citet{Gandhietal2019} estimated the distance of this source to be less 
than or around 1~kpc from Gaia DR2 observation.
The DR2 distance would make the source under-luminous in both optical and X-ray wavebands. Apart from that, a large discrepancy
also makes a large uncertainty in estimating the donor star classification and its intrinsic brightness.
This candidate showed two confirmed outbursts: one in 1987 \citep{Makino1987}, when the source entered into the soft state \citep{Kitamotoetal1990} and in 1997 \citep{Brocksoppetal2001},
when the source was in a pure hard state \citep{Revnivtsevetal2000}. In the same outburst for the first time, 
\citet{Fender1997} observed a radio counterpart. In late May 2015, the optical brightness of this source was 
twice higher than the quiescent values reported by \citet{RussellLewis2015}. 
Recently, \citet{Koljonen2016} showed that the
source has again entered into the hard state, 
studying multi-wavelength observation during its 2015 outburst. 
They also confirmed that optical/UV are tightly correlated 
with X-ray, which is consistent with the emission from the jet. 

\begin{figure}
\centering{
	\includegraphics[height=8.5truecm]{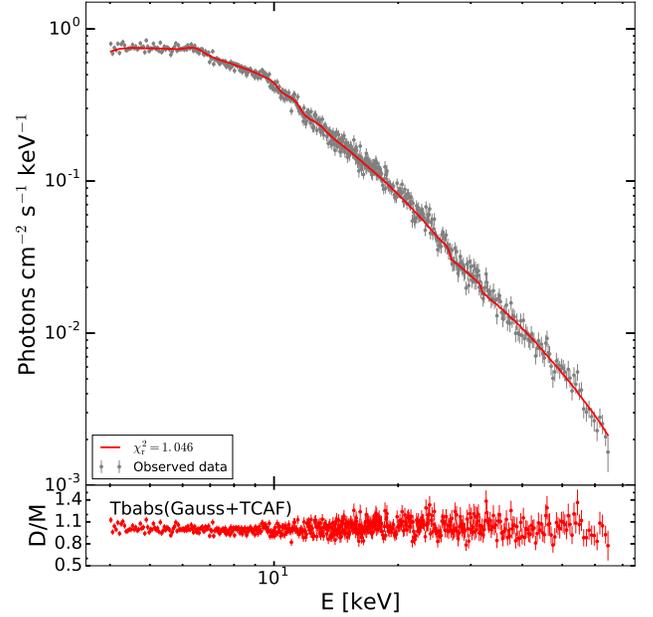}} 
	\caption{{\it NuSTAR} FPMA data (X02) of GS~1354-64, fitted with {\fontfamily{qcr}\selectfont TCAF} model as a local additive table model. Gray points with error 
	bar shows the observed data and red line is the model spectrum. Data/model (D/M) ratio is shown in the lower panel. 
	Data are rebinned for clarity.
}
\label{fig:ObsX02}
\end{figure}

We consider the {\it NuSTAR} \citep{Harrisonetal2013} observation of 2015 June 13 (90101006002, hereafter X02) and July 11 (X04) outbursts. 
We consider FPMA observation with 4.0-65.0~keV energy range data. The exposure times for these observations are $\sim24$ks
and $\sim 30$ks respectively. For pipeline reduction and spectral file generation we use {\fontfamily{qcr}\selectfont nupipelne} and {\fontfamily{qcr}\selectfont nuproducts} tasks. 
After successfully processing the data we do spectral fitting using (1) {\fontfamily{qcr}\selectfont Tbabs*(Gauss}+{\fontfamily{qcr}\selectfont TCAF)}, and (2) {\fontfamily{qcr}\selectfont Tabs*(Gauss}+{\fontfamily{qcr}\selectfont JeTCAF)}. 
We follow the standard data extraction procedures using NuSTARDAS and HEASoft packages and the spectral fitting using XSPEC. The detail of 
the analysis procedure and the related fitting issues are discussed in \citet{MondalChakrabarti2019}. For the spectral fitting of X02, we use {\fontfamily{qcr}\selectfont TCAF} model generated  
{\it fits} file \citep{Debnathetal2014} as a local additive table model, following {\fontfamily{qcr}\selectfont atable} command in {\fontfamily{qcr}\selectfont XSPEC}. For the generation of model fit file, large no of spectra are generated for the range between $10^{14}-10^{21}$ Hz.
During the fitting, we keep the black hole mass as a free parameter and for both the observations we keep $N_H$ fixed at $0.5\times10^{22} {\rm cm}^{-2}$ following HEASARC 
 column density tool \citep{DickeyLockman1990}. Data fitting with {\fontfamily{qcr}\selectfont TCAF} gives a very good fit with reduced $\chi^2=1.046$. 
 After successful fitting, we get the value of $M_{\rm BH}=7.22^{+0.25}_{-0.89}$. In \autoref{fig:ObsX02}, we show the {\fontfamily{qcr}\selectfont TCAF} model fitted spectra of X02. 

For the fitting of X04, first, we attempt to fit the data using {\fontfamily{qcr}\selectfont TCAF} model generated {\it fits} 
file. We tried with different combinations of model parameters, however, the quality of the 
fit is poor, with a reduced $\chi^2$ value $> 2.5$, as that version of the {\it fits} 
file does not include the jet contribution. After this, we fit the data with {\fontfamily{qcr}\selectfont JeTCAF} model keeping all parameters free provided the $N_{\rm H}$.
We see that {\fontfamily{qcr}\selectfont JeTCAF} model fits the observed data with a much better statistics. 
In this case, we run the {\it source code} directly in {\fontfamily{qcr}\selectfont XSPEC} as a table model using {\fontfamily{qcr}\selectfont initpackage} and
{\fontfamily{qcr}\selectfont lmod} tools \citep{Arnaud1996}. The model parameters range used in {\fontfamily{qcr}\selectfont lmod} to run {\fontfamily{qcr}\selectfont JeTCAF} in {\fontfamily{qcr}\selectfont XSPEC} is shown in Table~1. The range of energy considered for the generation of model spectrum to fit the data is between $10^{17}-10^{20}$ Hz. For both the models we assumed accretion efficiency 0.1.

\begin{figure}
\centering{
\includegraphics[height=8.5truecm]{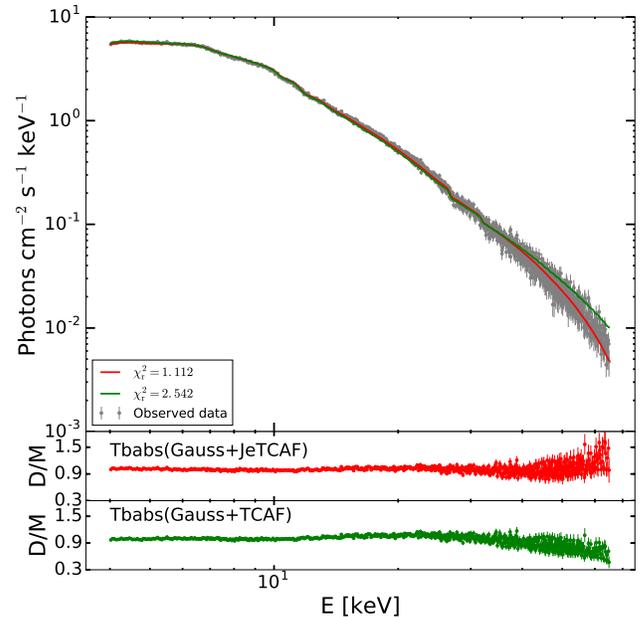}}
\caption{{\it NuSTAR} FPMA data (X04) of GS~1354-64, fitted with {\fontfamily{qcr}\selectfont JeTCAF} model as a local model directly in XSPEC. Gray points with error bars 
show the observed data and the red line is the model spectrum. The green line shows the {\fontfamily{qcr}\selectfont TCAF} model spectrum for comparison with {\fontfamily{qcr}\selectfont JeTCAF} model. The difference can be noted between $\sim 15-25$ keV and  near the tail of the spectrum due to  BMC of the photons scattered off the base of the jet.
Data/model (D/M) ratio is shown in lower panels. 
	Data are rebinned for clarity.
}
\label{fig:ObsX04}
\end{figure}

In \autoref{fig:ObsX04}, we show the {\fontfamily{qcr}\selectfont JeTCAF} model fitted spectra. In the upper panel, the red curve shows the  {\fontfamily{qcr}\selectfont JeTCAF} model spectrum and the gray dots with 
error bars show the observed spectra. In the lower panel, we show the ratio of data and model, which shows a good
agreement of fit with a reduced $\chi^2=1.112$. For comparison with {\fontfamily{qcr}\selectfont TCAF} model fit to the data, we overplotted the model spectrum (green line) and corresponding D/M ratio (in green) in the bottom panel. The model fitted parameters are shown in \autoref{table:results} for both X02 and X04 observations. The model fitted mass of the black hole for X04 is $6.7^{+0.54}_{-0.68}$.
Comparing mass obtained from both the observations, we conclude that the mass of the central black hole is $\sim 7\,M_\odot$.
After reaching a good fit, we use the model fitted parameters to recalculate the energy spectral index ($\alpha$) for X02 and X04. 
The calculated value of $\alpha$ for X02 and X04 are $0.63$ and $0.87$ respectively. The values of $\alpha$ and accretion rates 
confirm that on X02 day the source was in the ``hard spectral state'' and on X04 day it was in the ``intermediate spectral state'' 
when the jet was observed. The ratio of outflowing and inflowing matter rate for this observation is $0.12^{+0.02}_{-0.03}$, recalculated 
from the model fitted parameters using Eq.~1. The estimation of error in $R_{\dot m}$ measurement is discussed in Appendix. 
The above value of mass loss rate and the values of $R$ during the outburst, 
which is common in centrifugally driven flows \citep[][C99]{Moltenietal1994},
show the strong signature of an outflowing jet. Our model fitted mass is well within the range reported earlier for this source. 

\begin{table} 
\centering
\caption{Best fitted \label{table:results} {\fontfamily{qcr}\selectfont Gauss+TCAF} and {\fontfamily{qcr}\selectfont Gauss+JeTCAF} model fitted parameters. 
The Gaussian model fitted line energy is $E_{\rm g}$ and width is $\sigma_{\rm g}$. Here, $N_{\rm g}$ is the {\fontfamily{qcr}\selectfont Gaussian} model normalization in Photons\,cm$^{-2}$\,s$^{-1}$ unit.}
\resizebox{8cm}{!}{
\begin{tabular}{l c r}
\hline
&{\fontfamily{qcr}\selectfont TCAF}    & {\fontfamily{qcr}\selectfont JeTCAF} \\
Paramaters&           X02             &            X04\\
\hline
$E_{\rm g}$ [keV]&$6.40^{+0.06}_{-0.06}$ & $6.45^{+0.05}_{-0.06}$\\
$\sigma_{\rm g}$ [keV]&$0.24^{+0.06}_{-0.07}$ &$0.60^{+0.04}_{-0.04}$ \\
$N_{\rm g}[10^{-4}]$&$1.41^{+0.22}_{-0.25}$ &$17.28^{+1.17}_{-1.05}$ \\
$M_{\rm BH}$ [$M_\odot$]&$7.22^{+0.25}_{-0.89}$   & $6.70^{+0.54}_{-0.68}$ \\
$\dot{m}_{\rm d}$ [$\dot M_{\rm Edd}$]&$0.038^{+0.005}_{-0.003}$  & $0.084^{+0.003}_{-0.003}$  \\
$\dot{m}_{\rm h}$ [$\dot M_{\rm Edd}$]&$1.346^{+0.092}_{-0.076}$ & $0.265^{+0.009}_{-0.013}$  \\
$X_{\rm s}$ $[r_{\rm g}$]&$140.24^{+18.07}_{-5.89}$       & $37.74^{+3.49}_{-2.74}$   \\
$R$&$1.94^{+0.13}_{-0.18}$            & $4.22^{+0.19}_{-0.28}$   \\
$f_{\rm col}$& $-$                    & $0.47^{+0.09}_{-0.09}$  \\
\hline
&$\chi_{\rm r}^2=1.046$    & $\chi_{\rm r}^2=1.112$ \\
\hline
\end{tabular}
}    
\end{table}

Earlier reports confirmed this outburst as a ``failed outburst'' due to the presence of only hard state \citep{ElBataletal2016,StieleKong2016}. 
From our model fitted parameters, $\alpha$ value and high mass outflow rate, we 
identify the spectral state to be a hard (intermediate) state, like earlier findings. It is to be noticed that disk accretion rate increased 
by $\sim 2.2$~times, and entered into the region from where catastrophic cooling becomes significant according to {\fontfamily{qcr}\selectfont TCAF}.
As we discussed earlier, from the model fitted parameters, mass loss rate, and energy spectral index, we confirm the 
presence of both hard and intermediate states.

This candidate showed a low-frequency QPO ($\sim 0.2$~Hz) with increasing centroid frequency \citep{Koljonen2016}, which might 
be the indication of some instability occurring periodically. From the high flux behavior several authors argued this source 
to be similar to GRS~1915+105. Another possibility could be a periodic 
feedback from jet \citep{Chakrabartietal2002,Vadawaleetal2001}. 
There is no detailed dynamical 
lightcurve analysis on this topic till date. If we consider the burst time ($t_{\rm b-b}$) theory of C99 and 
estimate the periodicity of the burst from the model fitted parameters, it gives $\sim 20$~sec (see Eq.~24 of C99). The $t_{\rm b-b}$ value can also be estimated from the observed QPO frequency for this source (0.2\,Hz), using (C99) $t_{\rm b-b} \propto \nu_{\rm QPO}^{-2}$ $\sim 25$ s. Both estimates are consistent with each other. This estimate also matches with 20\,s periodicity observed in the lightcurve of this source during {\it RXTE} era \citep[see Fig. 2 of][]{JaniukBozena2011}.
There are several works in the literature \citep{Bellonietal2000,Raoetal2000,Naiketal2002,Chakrabartietal2004,PalChakrabarti2015} which discussed changes in different classes of lightcurves
in black hole candidates in a matter of few seconds. Such quick transitions are possible only if the
changes in accretion/ejection processes are due to local effects at the base of the jet or due to the return flow onto the disk \citep[e.g., see][]{Chakrabartietal2002}, whereas burst-on and burst-off states of GRS~1915+105 were discussed as due to interacting outflow with the
inner disk. Here we suggested cycles inside a lightcurve due to the feedback effects from the outflows in
a timescale of 20~sec. It is possible and is seen in GRS~1915+105 and IGR~17091-3624 \citep{Bellonietal2000,Rothsteinetal2005,PalChakrabarti2015} 
very often. There has to be changes in the local disk rather than the global disk. 

\section{Summary and Conclusions} \label{sec:Conclusion}
In this paper, we study the spectral properties of the accretion flows around black holes when 
the jet is also included in the {\fontfamily{qcr}\selectfont TCAF} solution. We apply this {\fontfamily{qcr}\selectfont JeTCAF} solution to study the transient 
source GS~1354-64 during its 2015 outburst using the data from {\it NuSTAR} satellite. We treat the base of 
the jet to be an additional Compton cloud component which also scatters soft photons from the disk. We also include
the effects of bulk motion of the diverging jet. Our jet configuration is self-consistent with the accretion dynamics in the sense that it estimates mass loss solving flow equations, rather than considering it separately. In this model, 
the jet is originated from the post-shock region (CENBOL) of the flow, and the velocity 
of the outflowing matter is calculated from the average temperature of the CENBOL
We estimate the mass outflow rate from the inflow rate. We use a realistic outflow 
to obtain the optical depth and the temperature of the Compton cloud. 
The mass outflow rate and the velocity are high in hard and intermediate states, thus the 
jet contribution to the spectrum is also modified significantly (due to bulk motion) during that time. In the soft state, the temperature is not high enough
to accelerate the mass outflow. Thus, the spectrum remains nearly blackbody. We also see that the contribution of the jet to the 
overall spectrum varies with the size of the Compton cloud, the shock compression ratio and the jet collimation factor. The overall fits using our simple consideration is acceptable, and we do not require presence of magnetic field or synchrotron emission to fit the data.

In this manuscript, we applied the model on the BHC GS\,1354-64 and calculated the mass outflow rate during
its 2015 outburst. In future, we aim to consider other jetted candidates using the same model. 
Our study shows that during the outburst, this candidate had a ratio of mass outflow to mass inflow rate of about $0.12^{+0.02}_{-0.03}$.
We suspect the reason behind the presence of very low QPO frequency with varying centroid frequency is also an indication of 
micro outbursts or feedback from jet to the disk occurring in very small-time scales (20~sec). This is similar to 
what happened in GRS~1915+105 (also in IGR~17091-3624). It is true that the spin of the black hole is important for both the disk and jet study, however its effects cannot spread beyond a few gravitational 
radii, while the physics we discuss is of farther out regions. 
As we discussed earlier, the spin may or may not contribute in powering the jet. 
However, the present {\fontfamily{qcr}\selectfont TCAF} or {\fontfamily{qcr}\selectfont JeTCAF} model does not consider the spin effect. In future, we aim to implement that effect self-consistently in the model.
A further study of this object is required to understand the jet feedback mechanism, class
transition in lightcurve to understand the dynamics of the disk.
 
\section{Acknowledgment}
We thank the referee for making insightful comments and suggestions. SM is thankful to Keith A. Arnaud for helping in model inclusion during his visit to NASA/GSFC as a student of 
COSPAR Capacity-building Workshop Fellowship program jointly with ISRO, India.
SM acknowledges Ramanujan Fellowship (\# RJF/2020/000113) by SERB-DST, Govt. of India for this research. 
This research has made use of the {\it NuSTAR} Data Analysis Software (NuSTARDAS) jointly developed by the ASI Science
Data Center (ASDC, Italy) and the California Institute of Technology (Caltech, USA).

\section*{APPENDIX} \label{sec:Appendix}
For the estimation of error in $R_{\dot m}$, we derived the relation:

$$\sigma_{R_{\dot m}}=\sqrt{ \left(\frac{\partial R_{\dot m}}{\partial R}\right)^2 \sigma_R^2 + \left(\frac{\partial R_{\dot m}}{\partial f_{\rm col}}\right)^2 \sigma_{fcol}^2},$$
where, $\sigma_R$ and $\sigma_{f_{\rm col}}$ are the error in measurement of shock compression ratio (R) and collimation factor ($f_{\rm col}$) respectively.
After doing a few steps of algebra we get the value of $\frac{\partial R_{\dot m}}{\partial R}$=-0.05 and $\frac{\partial R_{\dot m}}{\partial f_{\rm col}}$=0.24, which give the value of $\sigma_{R_{\dot m}}$ = \{+0.02, -0.03\}.



\begin{thebibliography}{}
\expandafter\ifx\csname natexlab\endcsname\relax\def\natexlab#1{#1}\fi
\providecommand{\url}[1]{\href{#1}{#1}}
\providecommand{\dodoi}[1]{doi:~\href{http://doi.org/#1}{\nolinkurl{#1}}}
\providecommand{\doeprint}[1]{\href{http://ascl.net/#1}{\nolinkurl{http://ascl.net/#1}}}
\providecommand{\doarXiv}[1]{\href{https://arxiv.org/abs/#1}{\nolinkurl{https://arxiv.org/abs/#1}}}

\bibitem[{{Arnaud}(1996)}]{Arnaud1996}
{Arnaud}, K.~A. 1996, Astronomical Society of the Pacific Conference Series,
  Vol. 101, {XSPEC: The First Ten Years}, ed. G.~H. {Jacoby} \& J.~{Barnes}, 17

\bibitem[{{Belloni} {et~al.}(2005){Belloni}, {Homan}, {Casella}, {van der
  Klis}, {Nespoli}, {Lewin}, {Miller}, \& {M{\'e}ndez}}]{Bellonietal2005}
{Belloni}, T., {Homan}, J., {Casella}, P., {et~al.} 2005, \aap, 440, 207,
  \dodoi{10.1051/0004-6361:20042457}

\bibitem[{{Belloni} {et~al.}(2000){Belloni}, {Migliari}, \&
  {Fender}}]{Bellonietal2000}
{Belloni}, T., {Migliari}, S., \& {Fender}, R.~P. 2000, \aap, 358, L29.
\newblock \doarXiv{astro-ph/0005307}

\bibitem[{{Blandford} {et~al.}(2019){Blandford}, {Meier}, \&
  {Readhead}}]{Blandfordetal2019}
{Blandford}, R., {Meier}, D., \& {Readhead}, A. 2019, \araa, 57, 467,
  \dodoi{10.1146/annurev-astro-081817-051948}

\bibitem[{{Blandford} \& {Payne}(1981)}]{BlandfordPayne1981}
{Blandford}, R.~D., \& {Payne}, D.~G. 1981, \mnras, 194, 1033,
  \dodoi{10.1093/mnras/194.4.1033}

\bibitem[{{Blandford} \& {Payne}(1982)}]{BlandfordPayne1982}
---. 1982, \mnras, 199, 883, \dodoi{10.1093/mnras/199.4.883}

\bibitem[{{Brocksopp} {et~al.}(2001){Brocksopp}, {Jonker}, {Fender}, {Groot},
  {van der Klis}, \& {Tingay}}]{Brocksoppetal2001}
{Brocksopp}, C., {Jonker}, P.~G., {Fender}, R.~P., {et~al.} 2001, \mnras, 323,
  517, \dodoi{10.1046/j.1365-8711.2001.04193.x}

\bibitem[{{Bromberg} {et~al.}(2011){Bromberg}, {Nakar}, {Piran}, \&
  {Sari}}]{Bromberg2011}
{Bromberg}, O., {Nakar}, E., {Piran}, T., \& {Sari}, R. 2011, \apj, 740, 100,
  \dodoi{10.1088/0004-637X/740/2/100}

\bibitem[{{Casares} {et~al.}(2004){Casares}, {Zurita}, {Shahbaz}, {Charles}, \&
  {Fender}}]{Casaresetal2004}
{Casares}, J., {Zurita}, C., {Shahbaz}, T., {Charles}, P.~A., \& {Fender},
  R.~P. 2004, \apjl, 613, L133, \dodoi{10.1086/425145}

\bibitem[{{Casares} {et~al.}(2009){Casares}, {Orosz}, {Zurita}, {Shahbaz},
  {Corral-Santana}, {McClintock}, {Garcia}, {Mart{\'\i}nez-Pais}, {Charles},
  {Fender}, \& {Remillard}}]{Casaresetal2009}
{Casares}, J., {Orosz}, J.~A., {Zurita}, C., {et~al.} 2009, \apjs, 181, 238,
  \dodoi{10.1088/0067-0049/181/1/238}

\bibitem[{{Chakrabarti} \& {Titarchuk}(1995)}]{ChakrabartiTitarchuk1995}
{Chakrabarti}, S., \& {Titarchuk}, L.~G. 1995, \apj, 455, 623,
  \dodoi{10.1086/176610}

\bibitem[{{Chakrabarti}(1989)}]{Chakrabarti1989}
{Chakrabarti}, S.~K. 1989, \apj, 347, 365, \dodoi{10.1086/168125}

\bibitem[{{Chakrabarti}(1996)}]{Chakrabarti1996}
---. 1996, \apj, 464, 664, \dodoi{10.1086/177354}

\bibitem[{{Chakrabarti}(1997)}]{Chakrabarti1997}
---. 1997, \apj, 484, 313, \dodoi{10.1086/304325}

\bibitem[{{Chakrabarti}(1999)}]{Chakrabarti99}
---. 1999, \aap, 351, 185.
\newblock \doarXiv{astro-ph/9910014}

\bibitem[{{Chakrabarti} {et~al.}(2004){Chakrabarti}, {Nandi}, {Choudhury}, \&
  {Chatterjee}}]{Chakrabartietal2004}
{Chakrabarti}, S.~K., {Nandi}, A., {Choudhury}, A., \& {Chatterjee}, U. 2004,
  \apj, 607, 406, \dodoi{10.1086/383235}

\bibitem[{{Chakrabarti} {et~al.}(2002){Chakrabarti}, {Nandi}, {Manickam},
  {Mandal}, \& {Rao}}]{Chakrabartietal2002}
{Chakrabarti}, S.~K., {Nandi}, A., {Manickam}, S.~G., {Mandal}, S., \& {Rao},
  A.~R. 2002, \apjl, 579, L21, \dodoi{10.1086/344783}

\bibitem[{{Chatterjee} {et~al.}(2016){Chatterjee}, {Debnath}, {Chakrabarti},
  {Mondal}, \& {Jana}}]{Chatterjeetal2016}
{Chatterjee}, D., {Debnath}, D., {Chakrabarti}, S.~K., {Mondal}, S., \& {Jana},
  A. 2016, \apj, 827, 88, \dodoi{10.3847/0004-637X/827/1/88}

\bibitem[{{Chattopadhyay} {et~al.}(2004){Chattopadhyay}, {Das}, \&
  {Chakrabarti}}]{Chattopadhyayetal2004}
{Chattopadhyay}, I., {Das}, S., \& {Chakrabarti}, S.~K. 2004, \mnras, 348, 846,
  \dodoi{10.1111/j.1365-2966.2004.07398.x}

\bibitem[{{Debnath} {et~al.}(2014){Debnath}, {Chakrabarti}, \&
  {Mondal}}]{Debnathetal2014}
{Debnath}, D., {Chakrabarti}, S.~K., \& {Mondal}, S. 2014, \mnras, 440, L121,
  \dodoi{10.1093/mnrasl/slu024}

\bibitem[{{Debnath} {et~al.}(2015){Debnath}, {Mondal}, \&
  {Chakrabarti}}]{Debnathetal2015}
{Debnath}, D., {Mondal}, S., \& {Chakrabarti}, S.~K. 2015, \mnras, 447, 1984,
  \dodoi{10.1093/mnras/stu2588}

\bibitem[{{Dickey} \& {Lockman}(1990)}]{DickeyLockman1990}
{Dickey}, J.~M., \& {Lockman}, F.~J. 1990, \araa, 28, 215,
  \dodoi{10.1146/annurev.aa.28.090190.001243}

\bibitem[{{Done} {et~al.}(2012){Done}, {Davis}, {Jin}, {Blaes}, \&
  {Ward}}]{Doneetal2012}
{Done}, C., {Davis}, S.~W., {Jin}, C., {Blaes}, O., \& {Ward}, M. 2012, \mnras,
  420, 1848, \dodoi{10.1111/j.1365-2966.2011.19779.x}

\bibitem[{{Done} {et~al.}(2007){Done}, {Gierli{\'n}ski}, \&
  {Kubota}}]{Doneetal2007}
{Done}, C., {Gierli{\'n}ski}, M., \& {Kubota}, A. 2007, \aapr, 15, 1,
  \dodoi{10.1007/s00159-007-0006-1}

\bibitem[{{D'Silva} \& {Chakrabarti}(1994)}]{DsilvaChak1994}
{D'Silva}, S., \& {Chakrabarti}, S.~K. 1994, \apj, 424, 149,
  \dodoi{10.1086/173879}

\bibitem[{{Ebisawa} {et~al.}(2003){Ebisawa}, {{\.Z}ycki}, {Kubota}, {Mizuno},
  \& {Watarai}}]{Ebisawaetal2003}
{Ebisawa}, K., {{\.Z}ycki}, P., {Kubota}, A., {Mizuno}, T., \& {Watarai}, K.-y.
  2003, \apj, 597, 780, \dodoi{10.1086/378586}

\bibitem[{{Eggum} {et~al.}(1988){Eggum}, {Coroniti}, \& {Katz}}]{Eggumetal1988}
{Eggum}, G.~E., {Coroniti}, F.~V., \& {Katz}, J.~I. 1988, \apj, 330, 142,
  \dodoi{10.1086/166462}

\bibitem[{{El-Batal} {et~al.}(2016){El-Batal}, {Miller}, {Reynolds}, {Boggs},
  {Chistensen}, {Craig}, {Fuerst}, {Hailey}, {Harrison}, {Stern}, {Tomsick},
  {Walton}, \& {Zhang}}]{ElBataletal2016}
{El-Batal}, A.~M., {Miller}, J.~M., {Reynolds}, M.~T., {et~al.} 2016, \apjl,
  826, L12, \dodoi{10.3847/2041-8205/826/1/L12}

\bibitem[{{Fender} {et~al.}(2004){Fender}, {Belloni}, \&
  {Gallo}}]{Fenderetal2004}
{Fender}, R.~P., {Belloni}, T.~M., \& {Gallo}, E. 2004, \mnras, 355, 1105,
  \dodoi{10.1111/j.1365-2966.2004.08384.x}

\bibitem[{{Fender} {et~al.}(2010){Fender}, {Gallo}, \& {Russell}}]{Fender2010}
{Fender}, R.~P., {Gallo}, E., \& {Russell}, D. 2010, \mnras, 406, 1425,
  \dodoi{10.1111/j.1365-2966.2010.16754.x}

\bibitem[{{Fender} {et~al.}(2009){Fender}, {Homan}, \&
  {Belloni}}]{Fenderatal2009}
{Fender}, R.~P., {Homan}, J., \& {Belloni}, T.~M. 2009, \mnras, 396, 1370,
  \dodoi{10.1111/j.1365-2966.2009.14841.x}

\bibitem[{{Fender} {et~al.}(1997){Fender}, {Tingay}, {Higdon}, {Wark}, \&
  {Wieringa}}]{Fender1997}
{Fender}, R.~P., {Tingay}, S.~J., {Higdon}, J., {Wark}, R., \& {Wieringa}, M.
  1997, \iaucirc, 6779, 2

\bibitem[{{Gandhi} {et~al.}(2019){Gandhi}, {Rao}, {Johnson}, {Paice}, \&
  {Maccarone}}]{Gandhietal2019}
{Gandhi}, P., {Rao}, A., {Johnson}, M. A.~C., {Paice}, J.~A., \& {Maccarone},
  T.~J. 2019, \mnras, 485, 2642, \dodoi{10.1093/mnras/stz438}

\bibitem[{{Garain} {et~al.}(2020){Garain}, {Balsara}, {Chakrabarti}, \&
  {Kim}}]{Garainetal2020}
{Garain}, S.~K., {Balsara}, D.~S., {Chakrabarti}, S.~K., \& {Kim}, J. 2020,
  \apj, 888, 59, \dodoi{10.3847/1538-4357/ab5d3c}

\bibitem[{{Garain} {et~al.}(2012){Garain}, {Ghosh}, \&
  {Chakrabarti}}]{Garainetal2012}
{Garain}, S.~K., {Ghosh}, H., \& {Chakrabarti}, S.~K. 2012, \apj, 758, 114,
  \dodoi{10.1088/0004-637X/758/2/114}

\bibitem[{{Harrison} {et~al.}(2013){Harrison}, {Craig}, {Christensen},
  {Hailey}, {Zhang}, {Boggs}, {Stern}, {Cook}, {Forster}, {Giommi},
  {Grefenstette}, {Kim}, {Kitaguchi}, {Koglin}, {Madsen}, {Mao}, {Miyasaka},
  {Mori}, {Perri}, {Pivovaroff}, {Puccetti}, {Rana}, {Westergaard}, {Willis},
  {Zoglauer}, {An}, {Bachetti}, {Barri{\`e}re}, {Bellm}, {Bhalerao},
  {Brejnholt}, {Fuerst}, {Liebe}, {Markwardt}, {Nynka}, {Vogel}, {Walton},
  {Wik}, {Alexander}, {Cominsky}, {Hornschemeier}, {Hornstrup}, {Kaspi},
  {Madejski}, {Matt}, {Molendi}, {Smith}, {Tomsick}, {Ajello}, {Ballantyne},
  {Balokovi{\'c}}, {Barret}, {Bauer}, {Blandford}, {Brandt}, {Brenneman},
  {Chiang}, {Chakrabarty}, {Chenevez}, {Comastri}, {Dufour}, {Elvis}, {Fabian},
  {Farrah}, {Fryer}, {Gotthelf}, {Grindlay}, {Helfand}, {Krivonos}, {Meier},
  {Miller}, {Natalucci}, {Ogle}, {Ofek}, {Ptak}, {Reynolds}, {Rigby},
  {Tagliaferri}, {Thorsett}, {Treister}, \& {Urry}}]{Harrisonetal2013}
{Harrison}, F.~A., {Craig}, W.~W., {Christensen}, F.~E., {et~al.} 2013, \apj,
  770, 103, \dodoi{10.1088/0004-637X/770/2/103}

\bibitem[{{Hjellming} \& {Rupen}(1995)}]{Hjellming1995}
{Hjellming}, R.~M., \& {Rupen}, M.~P. 1995, \nat, 375, 464,
  \dodoi{10.1038/375464a0}

\bibitem[{{Jana} {et~al.}(2017){Jana}, {Chakrabarti}, \&
  {Debnath}}]{Janaetal2017}
{Jana}, A., {Chakrabarti}, S.~K., \& {Debnath}, D. 2017, \apj, 850, 91,
  \dodoi{10.3847/1538-4357/aa88a5}

\bibitem[{{Janiuk} \& {Czerny}(2011)}]{JaniukBozena2011}
{Janiuk}, A., \& {Czerny}, B. 2011, \mnras, 414, 2186,
  \dodoi{10.1111/j.1365-2966.2011.18544.x}

\bibitem[{{Kawashima} {et~al.}(2012){Kawashima}, {Ohsuga}, {Mineshige},
  {Yoshida}, {Heinzeller}, \& {Matsumoto}}]{Kawashimaetal2012}
{Kawashima}, T., {Ohsuga}, K., {Mineshige}, S., {et~al.} 2012, \apj, 752, 18,
  \dodoi{10.1088/0004-637X/752/1/18}

\bibitem[{{King} {et~al.}(2001){King}, {Davies}, {Ward}, {Fabbiano}, \&
  {Elvis}}]{Kingetal2001}
{King}, A.~R., {Davies}, M.~B., {Ward}, M.~J., {Fabbiano}, G., \& {Elvis}, M.
  2001, \apjl, 552, L109, \dodoi{10.1086/320343}

\bibitem[{{Kitamoto} {et~al.}(1990){Kitamoto}, {Tsunemi}, {Pedersen},
  {Ilovaisky}, \& {van der Klis}}]{Kitamotoetal1990}
{Kitamoto}, S., {Tsunemi}, H., {Pedersen}, H., {Ilovaisky}, S.~A., \& {van der
  Klis}, M. 1990, \apj, 361, 590, \dodoi{10.1086/169222}

\bibitem[{{Koljonen} {et~al.}(2016){Koljonen}, {Russell}, {Corral-Santana},
  {Armas Padilla}, {Mu{\~n}oz-Darias}, {Lewis}, {Coriat}, \&
  {Bauer}}]{Koljonen2016}
{Koljonen}, K.~I.~I., {Russell}, D.~M., {Corral-Santana}, J.~M., {et~al.} 2016,
  \mnras, 460, 942, \dodoi{10.1093/mnras/stw1007}

\bibitem[{{Makino}(1987)}]{Makino1987}
{Makino}, F. 1987, \iaucirc, 4342, 1

\bibitem[{{Miller-Jones} {et~al.}(2012){Miller-Jones}, {Sivakoff},
  {Altamirano}, {Coriat}, {Corbel}, {Dhawan}, {Krimm}, {Remillard}, {Rupen},
  {Russell}, {Fender}, {Heinz}, {K{\"o}rding}, {Maitra}, {Markoff}, {Migliari},
  {Sarazin}, \& {Tudose}}]{MillerJones2012}
{Miller-Jones}, J.~C.~A., {Sivakoff}, G.~R., {Altamirano}, D., {et~al.} 2012,
  \mnras, 421, 468, \dodoi{10.1111/j.1365-2966.2011.20326.x}

\bibitem[{{Mirabel} \& {Rodr{\'\i}guez}(1998)}]{Mirabel1995}
{Mirabel}, I.~F., \& {Rodr{\'\i}guez}, L.~F. 1998, \nat, 392, 673,
  \dodoi{10.1038/33603}

\bibitem[{{Molteni} {et~al.}(1994){Molteni}, {Lanzafame}, \&
  {Chakrabarti}}]{Moltenietal1994}
{Molteni}, D., {Lanzafame}, G., \& {Chakrabarti}, S.~K. 1994, \apj, 425, 161,
  \dodoi{10.1086/173972}

\bibitem[{{Mondal} \& {Chakrabarti}(2013)}]{MondalChakrabart2013}
{Mondal}, S., \& {Chakrabarti}, S.~K. 2013, \mnras, 431, 2716,
  \dodoi{10.1093/mnras/stt361}

\bibitem[{{Mondal} \& {Chakrabarti}(2019)}]{MondalChakrabarti2019}
---. 2019, \mnras, 483, 1178, \dodoi{10.1093/mnras/sty3169}

\bibitem[{{Mondal} {et~al.}(2014{\natexlab{a}}){Mondal}, {Chakrabarti}, \&
  {Debnath}}]{Mondaletal2014b}
{Mondal}, S., {Chakrabarti}, S.~K., \& {Debnath}, D. 2014{\natexlab{a}}, \apss,
  353, 223, \dodoi{10.1007/s10509-014-2008-6}

\bibitem[{{Mondal} {et~al.}(2017){Mondal}, {Chakrabarti}, {Nagarkoti}, \&
  {Ar{\'e}valo}}]{Mondaletal2017}
{Mondal}, S., {Chakrabarti}, S.~K., {Nagarkoti}, S., \& {Ar{\'e}valo}, P. 2017,
  \apj, 850, 47, \dodoi{10.3847/1538-4357/aa7e27}

\bibitem[{{Mondal} {et~al.}(2014{\natexlab{b}}){Mondal}, {Debnath}, \&
  {Chakrabarti}}]{Mondaletal2014}
{Mondal}, S., {Debnath}, D., \& {Chakrabarti}, S.~K. 2014{\natexlab{b}}, \apj,
  786, 4, \dodoi{10.1088/0004-637X/786/1/4}

\bibitem[{{Nagarkoti} \& {Chakrabarti}(2016)}]{NagarkotiChakrabarti2016}
{Nagarkoti}, S., \& {Chakrabarti}, S.~K. 2016, \mnras, 462, 850,
  \dodoi{10.1093/mnras/stw1700}

\bibitem[{{Naik} {et~al.}(2002){Naik}, {Rao}, \& {Chakrabarti}}]{Naiketal2002}
{Naik}, S., {Rao}, A.~R., \& {Chakrabarti}, S.~K. 2002, Journal of Astrophysics
  and Astronomy, 23, 213, \dodoi{10.1007/BF02702284}

\bibitem[{{Nandi} {et~al.}(2001){Nandi}, {Chakrabarti}, {Vadawale}, \&
  {Rao}}]{Nandietal2001}
{Nandi}, A., {Chakrabarti}, S.~K., {Vadawale}, S.~V., \& {Rao}, A.~R. 2001,
  \aap, 380, 245, \dodoi{10.1051/0004-6361:20011444}

\bibitem[{{Narayan} \& {McClintock}(2012)}]{Narayan2012}
{Narayan}, R., \& {McClintock}, J.~E. 2012, \mnras, 419, L69,
  \dodoi{10.1111/j.1745-3933.2011.01181.x}

\bibitem[{{Nobili} {et~al.}(1993){Nobili}, {Turolla}, \&
  {Zampieri}}]{Nobilietal1993}
{Nobili}, L., {Turolla}, R., \& {Zampieri}, L. 1993, \apj, 404, 686,
  \dodoi{10.1086/172322}

\bibitem[{{Ohsuga} {et~al.}(2005){Ohsuga}, {Mori}, {Nakamoto}, \&
  {Mineshige}}]{Ohsugaetal2005}
{Ohsuga}, K., {Mori}, M., {Nakamoto}, T., \& {Mineshige}, S. 2005, \apj, 628,
  368, \dodoi{10.1086/430728}

\bibitem[{{Okuda}(2002)}]{Okuda2002}
{Okuda}, T. 2002, \pasj, 54, 253, \dodoi{10.1093/pasj/54.2.253}

\bibitem[{{Pahari} {et~al.}(2017){Pahari}, {Gandhi}, {Charles}, {Kotze},
  {Altamirano}, \& {Misra}}]{Paharietal2017}
{Pahari}, M., {Gandhi}, P., {Charles}, P.~A., {et~al.} 2017, \mnras, 469, 193,
  \dodoi{10.1093/mnras/stx840}

\bibitem[{{Pal} \& {Chakrabarti}(2015)}]{PalChakrabarti2015}
{Pal}, P.~S., \& {Chakrabarti}, S.~K. 2015, Advances in Space Research, 56,
  1784, \dodoi{10.1016/j.asr.2015.07.016}

\bibitem[{{Radhika} \& {Nandi}(2014)}]{RadhikaNandi2014}
{Radhika}, D., \& {Nandi}, A. 2014, Advances in Space Research, 54, 1678,
  \dodoi{10.1016/j.asr.2014.06.039}

\bibitem[{{Rao} {et~al.}(2000){Rao}, {Yadav}, \& {Paul}}]{Raoetal2000}
{Rao}, A.~R., {Yadav}, J.~S., \& {Paul}, B. 2000, \apj, 544, 443,
  \dodoi{10.1086/317168}

\bibitem[{{Revnivtsev} {et~al.}(2000){Revnivtsev}, {Borozdin}, {Priedhorsky},
  \& {Vikhlinin}}]{Revnivtsevetal2000}
{Revnivtsev}, M.~G., {Borozdin}, K.~N., {Priedhorsky}, W.~C., \& {Vikhlinin},
  A. 2000, \apj, 530, 955, \dodoi{10.1086/308386}

\bibitem[{{Reynolds} \& {Miller}(2011)}]{ReynoldsMiller2011}
{Reynolds}, M.~T., \& {Miller}, J.~M. 2011, \apjl, 734, L17,
  \dodoi{10.1088/2041-8205/734/1/L17}

\bibitem[{{Rothstein} {et~al.}(2005){Rothstein}, {Eikenberry}, \&
  {Matthews}}]{Rothsteinetal2005}
{Rothstein}, D.~M., {Eikenberry}, S.~S., \& {Matthews}, K. 2005, \apj, 626,
  991, \dodoi{10.1086/429217}

\bibitem[{{Russell} \& {Lewis}(2015)}]{RussellLewis2015}
{Russell}, D.~M., \& {Lewis}, F. 2015, The Astronomer's Telegram, 7637, 1

\bibitem[{{Shakura} \& {Sunyaev}(1973)}]{ShakuraSunyaev1973}
{Shakura}, N.~I., \& {Sunyaev}, R.~A. 1973, \aap, 500, 33

\bibitem[{{Singh} \& {Chakrabarti}(2011)}]{SinghChakrabarti2011}
{Singh}, C.~B., \& {Chakrabarti}, S.~K. 2011, \mnras, 410, 2414,
  \dodoi{10.1111/j.1365-2966.2010.17615.x}

\bibitem[{{Sriram} {et~al.}(2012){Sriram}, {Rao}, \& {Choi}}]{Srirametal2012}
{Sriram}, K., {Rao}, A.~R., \& {Choi}, C.~S. 2012, \aap, 541, A6,
  \dodoi{10.1051/0004-6361/201218799}

\bibitem[{{Stiele} \& {Kong}(2016)}]{StieleKong2016}
{Stiele}, H., \& {Kong}, A.~K.~H. 2016, \mnras, 459, 4038,
  \dodoi{10.1093/mnras/stw903}

\bibitem[{{Sunyaev} \& {Titarchuk}(1980)}]{SunyaevTitarchuk1980}
{Sunyaev}, R.~A., \& {Titarchuk}, L.~G. 1980, \aap, 500, 167

\bibitem[{{Titarchuk} {et~al.}(2003){Titarchuk}, {Kazanas}, \&
  {Becker}}]{Titarchuketal2003}
{Titarchuk}, L., {Kazanas}, D., \& {Becker}, P.~A. 2003, \apj, 598, 411,
  \dodoi{10.1086/378701}

\bibitem[{{Titarchuk} \& {Shrader}(2005)}]{TitarchukShrader2005}
{Titarchuk}, L., \& {Shrader}, C. 2005, \apj, 623, 362, \dodoi{10.1086/424918}

\bibitem[{{Titarchuk} \& {Zannias}(1998)}]{TitarchukZannias1998}
{Titarchuk}, L., \& {Zannias}, T. 1998, \apj, 493, 863, \dodoi{10.1086/305157}

\bibitem[{{Vadawale} {et~al.}(2001){Vadawale}, {Rao}, {Nandi}, \&
  {Chakrabarti}}]{Vadawaleetal2001}
{Vadawale}, S.~V., {Rao}, A.~R., {Nandi}, A., \& {Chakrabarti}, S.~K. 2001,
  \aap, 370, L17, \dodoi{10.1051/0004-6361:20010318}

\end{thebibliography}



\end{document}